\documentclass[reprint,notitlepage,amsmath,amsfonts,amssymb,floatfix,nolongbibliography,superscriptaddress]{revtex4-2}
\usepackage[export]{adjustbox}
\usepackage{bm}
\usepackage{color}

\begin{document}
\title{Restoring permutational invariance in the Jordan-Wigner transformation}
\author{Thomas M. Henderson}
\affiliation{Department of Chemistry, Rice University, Houston, TX 77005-1892}
\affiliation{Department of Physics and Astronomy, Rice University, Houston, TX 77005-1892}

\author{Fei Gao}
\affiliation{Department of Physics and Astronomy, Rice University, Houston, TX 77005-1892}

\author{Gustavo E. Scuseria}
\affiliation{Department of Chemistry, Rice University, Houston, TX 77005-1892}
\affiliation{Department of Physics and Astronomy, Rice University, Houston, TX 77005-1892}
\date{\today}

\begin{abstract}
The Jordan-Wigner transformation is a powerful tool for converting systems of spins into systems of fermions, or vice versa.  While this mapping is exact, the transformation itself depends on the labeling of the spins.  One consequence of this dependence is that approximate solutions of a Jordan-Wigner--transformed Hamiltonian may depend on the (physically inconsequential) labeling of the spins.  In this work, we turn to an extended Jordan-Wigner transformation which remedies this problem and which may also introduce some correlation atop the Hartree-Fock solution of a transformed spin Hamiltonian.  We demonstrate that this extended Jordan-Wigner transformation can be thought of as arising from a unitary version of the Lie algebraic similarity transformation (LAST) theory.  We show how these ideas, particularly in combination with the standard (non-unitary) version of LAST, can provide a potentially powerful tool for the treatment of the XXZ and $J_1-J_2$ Heisenberg Hamiltonians.
\end{abstract}

\maketitle

\section{Introduction}
The difficulty in mathematically describing many physical systems depends on the framework one uses to treat them, so that a problem which is difficult from one perspective may be simple from another.  For example, constrained motion in classical mechanics is more easily described by using the Lagrangian formalism than by directly applying Newton's laws.

One manifestation of this phenomenon in quantum mechanics is the concept of duality: a system which is strongly correlated in one picture becomes weakly correlated in another.\cite{Batista2001}  The one-dimensional spin-1/2 XXZ Hamiltonian
\begin{equation}
H_{XXZ} = \sum_{\langle pq \rangle} \left[\frac{1}{2} \, \left(S_p^+ \, S_q^- + S_p^- \, S_q^+\right) + \Delta \, S_p^z \, S_q^z\right],
\label{Eqn:DefHXXZ}
\end{equation}
in which nearest-neighbor sites $p$ and $q$ have an anisotropic Heisenberg interaction, provides a textbook example of this kind of duality.\cite{Nishemori2011}  Although this problem can be somewhat tedious to solve in the language of spins, the Jordan-Wigner (JW) transformation\cite{Jordan1928}
\begin{subequations}
\label{Eqn:DefJW}
\begin{align}
S_p^+ &\mapsto c_p^\dagger \, \phi_p^\dagger,
\\
S_p^- &\mapsto c_p \, \phi_p,
\\
S_p^z &\mapsto \bar{n}_p = n_p - \frac{1}{2},
\\
\phi_p^\dagger &= \phi_p = \mathrm{e}^{\mathrm{i} \, \pi \, \sum_{k<p} n_k},
\label{Eqn:JWStrings}
\\
n_p &= c_p^\dagger \, c_p,
\end{align}
\end{subequations}
converts the spin Hamiltonian into a Hamiltonian of spinless fermions which, at $\Delta = 0$, is non-interacting and thus trivially solvable.

Recently, this result inspired us to examine the JW transformation as a tool for solving more general spin systems.\cite{Henderson2022}  In the nearest-neighbor XXZ model, the JW strings $\phi_p$ cancel out, accounting for the simplicity of the fermionic Hamiltonian at $\Delta = 0$.  This is not true in general, which perhaps explains why the JW transformation has not seen as much use in the description of spin systems as it could do.  However, when written in the language of fermions, the JW strings (which are many-body operators) act on single determinants as Thouless transformations\cite{Thouless1960} which one can readily handle.  Our results suggested that although the presence of these additional transformations complicates the implementation, one could obtain significantly improved results by using fermionic methods for the JW-transformed Hamiltonian in lieu of using equally expensive computational techniques on the underlying spin Hamiltonian.

In this paper, we explore two improvements upon our initial work.  First, as the reader may have noticed, the JW transformation depends on the order of the fermions because the summation in Eqn. \ref{Eqn:JWStrings} defining the JW string for site $p$ runs over sites $k < p$.  Although this dependence is of no real consequence when one solves the JW-transformed Hamiltonian exactly, approximate solutions will generally give different results for the same physical system with sites indexed in different orders.  We provide examples in Sec. \ref{Sec:OrderingDependence}, and wish to overcome this limitation.  Additionally, we seek to incorporate size-extensive correlation techniques.  Our previous work treated the fermionic JW-transformed Hamiltonian with Hartree-Fock followed by configuration interaction.  Hartree-Fock, when all symmetries are broken, is extensive,\cite{JimenezHoyos2011} meaning that the energy scales properly with system size, but the configuration interaction correction is not.\cite{SzaboOstlund,BartlettShavitt}  We would prefer to use standard coupled cluster theory,\cite{Paldus1999,Bartlett2007,BartlettShavitt} but this is precluded by the JW strings.  Here, we employ the Lie algebraic similarity transformation (LAST) theory\cite{Wahlen-Strothman2015} to add a thermodynamically extensive correlation correction to the Hartree-Fock treatment of the JW-transformed Hamiltonian.

\section{Ordering Dependence in the Jordan-Wigner Transformation
\label{Sec:OrderingDependence}}
To motivate our work, let us briefly examine the results of Hartree-Fock calculations on JW-transformed spin Hamiltonians for different orderings of the sites.

Consider, then, the one-dimensional (1D) XXZ model previously introduced.  The JW-transformed Hamiltonian is
\begin{equation}
H \mapsto \sum_{\langle pq \rangle} \left[\frac{1}{2} \, \left(c_p^\dagger \, \phi_p^\dagger \, \phi_q \, c_q + hc\right) + \Delta \, \bar{n}_p \, \bar{n}_q\right].
\end{equation}
The notation $\langle pq \rangle$ denotes that sites $p$ and $q$ are nearest neighbors.

\begin{figure*}[t]
\includegraphics[width=0.45\textwidth]{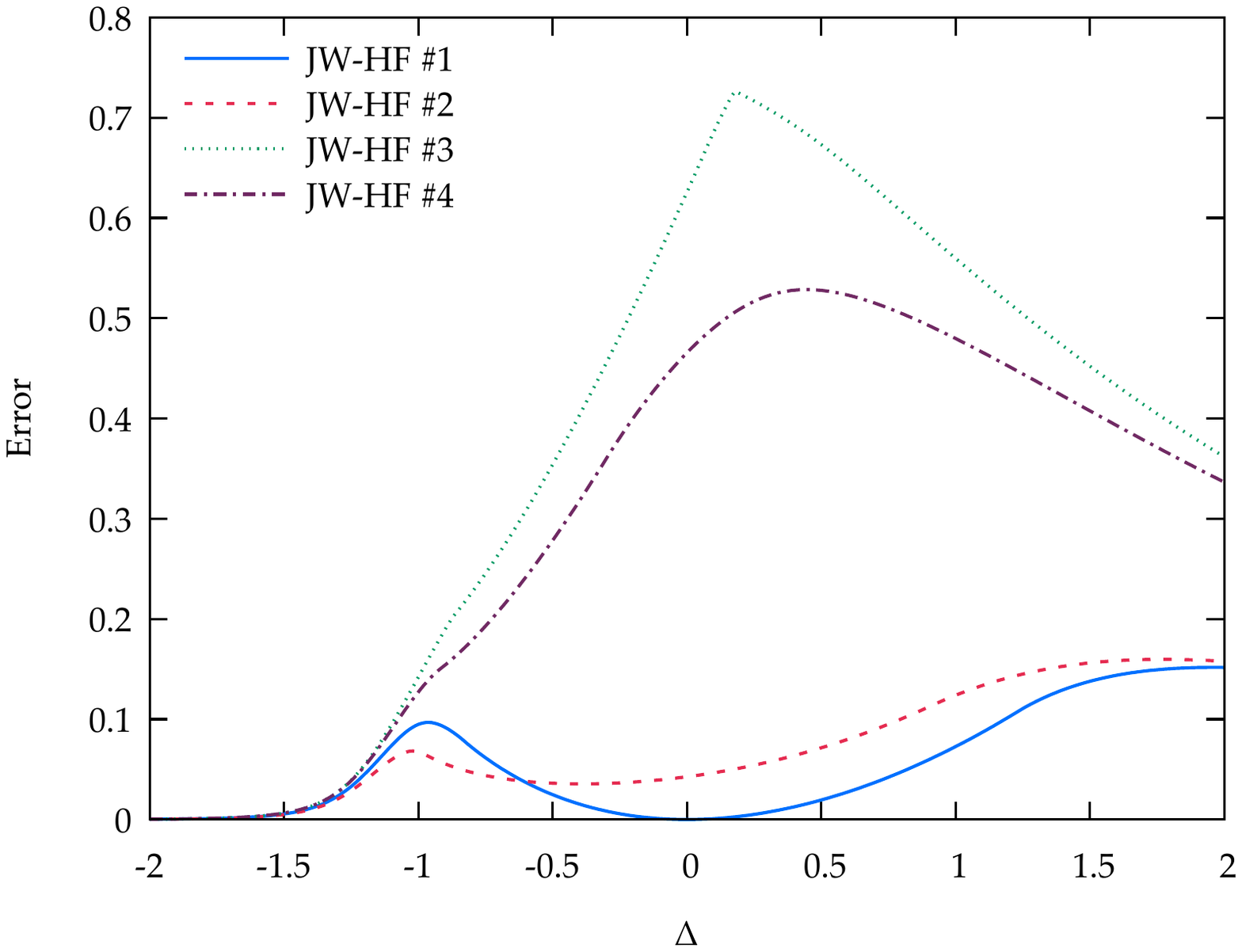}
\hfill
\includegraphics[width=0.45\textwidth]{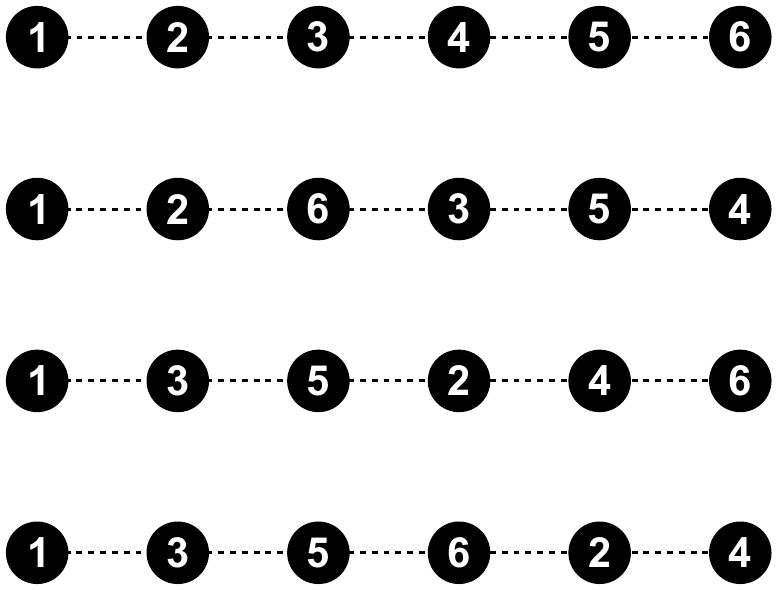}
\caption{Left panel: Hartree-Fock energy errors for different lattice labelings in the JW-transformed 6-site XXZ model with open boundary conditions and $S_z = 0$.  Right panel: Labeling schemes corresponding to the different curves, from \#1 to \#4, top to bottom.
\label{Fig:XXZOrderingHF}}
\end{figure*}

In a conventional ordering in which sites are numbered such that the nearest neighbors of site $p$ are sites $p \pm 1$, the JW strings cancel out.  For example,
\begin{subequations}
\begin{align}
c_p^\dagger \, \phi_p^\dagger \, \phi_{p+1} \, c_{p+1}
 &= c_p^\dagger \, \mathrm{e}^{\mathrm{i} \, \pi \, \sum_{k<p} n_k} \, \mathrm{e}^{\mathrm{i} \, \pi \, \sum_{k \le p} n_k} \, c_{p+1}
\\
 &= c_p^\dagger \, \mathrm{e}^{2 \, \mathrm{i} \, \pi \, \sum_{k<p} n_k} \, \mathrm{e}^{\mathrm{i} \, \pi \, n_p} \, c_{p+1}.
\end{align}
\end{subequations}
Using the fact that
\begin{equation}
\mathrm{e}^{\mathrm{i} \, \pi \, n_k} = 1 - 2 \, n_k
\end{equation}
so that
\begin{equation}
\mathrm{e}^{2 \, \mathrm{i} \, \pi \, n_k} = \left(1 - 2 \, n_k\right)^2 = 1 - 4 \, n_k + 4 \, n_k^2 = 1,
\end{equation}
we get
\begin{equation}
c_p^\dagger \, \phi_p^\dagger \, \phi_{p+1} \, c_{p+1} = c_p^\dagger \, \left(1 - 2 \, n_p\right) \, c_{p+1} = c_p^\dagger \, c_{p+1}
\end{equation}
since $c_p^\dagger \, n_p = c_p^\dagger \, c_p^\dagger \, c_p = 0$.  Thus, in this simple ordering, there are no JW strings to concern ourselves with, and
\begin{equation}
H \mapsto \sum_{\langle pq \rangle} \left[\frac{1}{2} \, \left(c_p^\dagger \, c_q + hc\right) + \Delta \, \bar{n}_p \, \bar{n}_q\right].
\label{Def:HXXZfermionic}
\end{equation}

On the other hand, there are $n!$ ways to label the $n$ sites, and only two labelings are such that the nearest neighbors of site $p$ are $p \pm 1$ for all $p$.  Different labelings produce different JW-transformed Hamiltonians, most of which have strings.  This means that in general we expect to get different results when we approximately solve a JW-transformed Hamiltonian with different labelings of the same underlying lattice.

This is not simply an academic concern. Figure \ref{Fig:XXZOrderingHF} demonstrates this phenomenon for the Hartree-Fock solutions of the JW-transformed 6-site 1D XXZ model with open boundary conditions and various labeling schemes.  Although relabeling the lattice sites does not affect the exact result, we can clearly obtain quite different numerical results by labeling the sites in different ways.  Fortunately, we can generalize the JW transformation in such a way as to eliminate dependence on the labeling of the sites.

\section{Extended Jordan-Wigner Transformations}
The Jordan-Wigner transformation as introduced in Eqn. \ref{Eqn:DefJW} is a special case of what we shall refer to as an extended Jordan-Wigner (EJW) transformation in which we write
\begin{subequations}
\label{Eqn:DefGeneralizedJW}
\begin{align}
S_p^+ &\mapsto c_p^\dagger \, \phi_p^\dagger,
\\
S_p^- &\mapsto c_p \, \phi_p,
\\
S_p^z &\mapsto \bar{n}_p,
\\
%\phi_p^\dagger &= \mathrm{e}^{\mathrm{i} \, \pi \, \sum_{q} \theta_{pq} \, n_q},
\phi_p^\dagger &= \mathrm{e}^{\mathrm{i} \, \sum_{q} \theta_{pq} \, n_q},
\end{align}
\end{subequations}
where the real parameters $\theta_{pq}$ satisfy the constraints
\begin{subequations}
\label{Eqn:ThetaConstraints}
\begin{align}
\theta_{pp} &= 0,
\\
|\theta_{pq} - \theta_{qp}| &= \pi.
\end{align}
\end{subequations}
A generic spin Hamiltonian $H_s$ of the form
\begin{align}
H_s = \sum_p H_p \, S_p^z + \sum_{p \ne q} \Big[&W_{pq} \, S_p^z \, S_q^z
\label{Eqn:DefHSpin}
\\
 &+\frac{1}{2} \, V_{pq} \, \left(S_p^+ \, S_q^- + S_q^+ \, S_p^-\right)\Big]
\nonumber
\end{align}
becomes, under this transformation, a $\boldsymbol{\theta}$-dependent fermionic Hamiltonian $H_f(\boldsymbol{\theta})$:
\begin{align}
H_f(\boldsymbol{\theta}) = \sum_p H_p \, \bar{n}_p &+ \sum_{p \ne q} \Big[W_{pq} \, \bar{n}_p \, \bar{n}_q
\\
 &+ \frac{1}{2} \, V_{pq} \, \left(c_p^\dagger \, \phi_p^\dagger \, \phi_q \, c_q + c_q^\dagger \, \phi_q^\dagger \, \phi_p \, c_p\right)\Big].
\nonumber
\end{align}

\begin{figure}[t]
\includegraphics[width=\columnwidth]{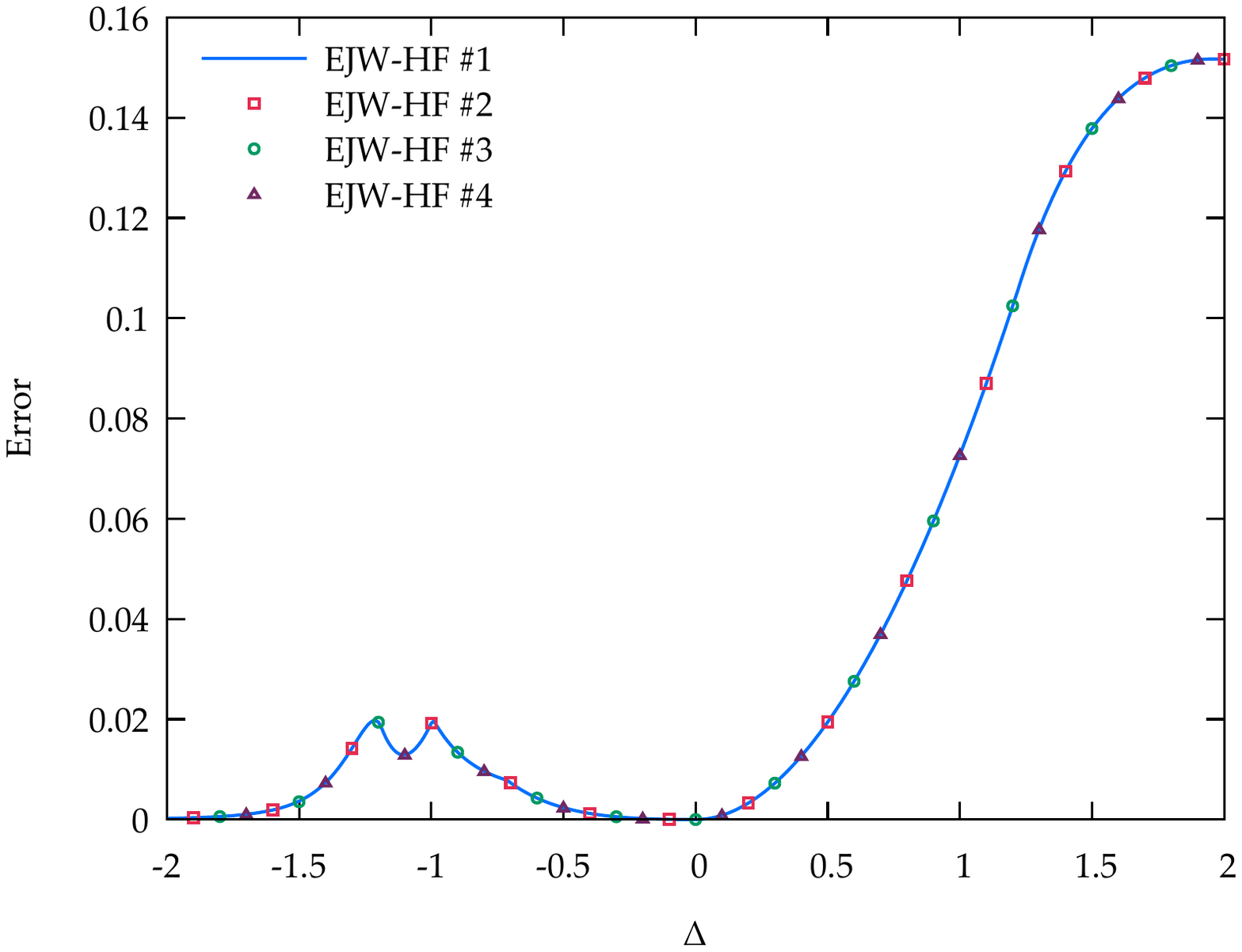}
\caption{Hartree-Fock energy errors for different lattice labelings in the EJW-transformed 6-site XXZ model with open boundary conditions and $S_z = 0$, using the same lattice labelings as in Fig. \ref{Fig:XXZOrderingHF}.
\label{Fig:XXZOrderingLAST}}
\end{figure}
This extended transformation has been mentioned in the literature,\cite{Wang1990} but to the best our knowledge has not been used in actual calculations.  Instead, $\boldsymbol{\theta}$ has been restricted to take on specific values.  For example, the Jordan-Wigner case is 
\begin{equation}
\theta_{pq}^{\mathrm{JW}} = 
\begin{cases}
0  &\qquad q>p,
\\
\pi &\qquad q<p.
\end{cases}
\end{equation}
In addition to the Jordan-Wigner limit, this extended Jordan-Wigner also encapsulates a lattice version of the Chern-Simons transformation (see, for example, Ref. \citenum{Yang2022}) in which one writes
\begin{equation}
\theta_{pq}^{\mathrm{CS}} = \mathrm{arg}(\vec{r}_p - \vec{r}_q)
\end{equation}
where $\vec{r}_p$, for example, is a vector pointing to site $p$.

We prefer a different approach.  Rather than choosing a specific form or specific values for the $\theta_{pq}$, we treat them as variational parameters in an optimization in which we solve the transformed Hamiltonian with Hartree-Fock theory.  To do so, we write the fermionic wave function as
\begin{equation}
|\mathrm{HF}(\boldsymbol{t})\rangle = \mathrm{e}^{\sum_{ia} t_i^a \, c_a^\dagger \, c_i} |0\rangle
\end{equation}
where $|0\rangle$ is some important lattice determinant; in the reference determinant $|0\rangle$, sites $i$ are occupied (corresponding to $\uparrow$ spin in the spin picture) and sites $a$ are unoccupied ($\downarrow$ spin in the spin picture).  The complex amplitudes $t_i^a$ are variational parameters specifying the Hartree-Fock wave function, and the Hartree-Fock energy
\begin{equation}
E_\mathrm{HF}(\boldsymbol{t},\boldsymbol{\theta}) = \frac{\langle \mathrm{HF}(\boldsymbol{t})| H_f(\boldsymbol{\theta}) |\mathrm{HF}(\boldsymbol{t})\rangle}{\langle \mathrm{HF}(\boldsymbol{t})|\mathrm{HF}(\boldsymbol{t})\rangle}
\end{equation}
depends on both the wave function amplitudes $\boldsymbol{t}$ and the Hamiltonian parameters $\boldsymbol{\theta}$ with respect to which we may then variationally minimize it.

It may not be entirely obvious, but this extended Jordan-Wigner transformation has an interesting consequence: the resulting energy is, if fully optimized, invariant to the ordering of the sites.  Figure \ref{Fig:XXZOrderingLAST} demonstrates this numerically.  We provide a detailed proof in Appendix \ref{Sec:AppendixGS}, but the substance of the proof is simple.  Changing the ordering of the sites permutes the Hamiltonian parameters $H_p$, $V_{pq}$, and $W_{pq}$ in a way which is not consequential but also permutes the rows and columns of the matrix $\boldsymbol{\theta}$.  We can always relabel sites after the transformation, and adjust values in the matrix $\boldsymbol{\theta}$, such that the fermionic Hamiltonian with sites expressed in the two different orders is the same.  In this we take advantage of the fact that adjusting a particular value $\theta_{pq} \to \theta_{pq} + 2 \, \pi$ is of no consequence because as we noted previously, $\mathrm{e}^{2 \, \pi \, \mathrm{i} \, n_q} = 1$.

Instead of combining Hartree-Fock theory with the EJW approach discussed here, one could imagine optimizing the ordering of the lattice sites with a standard JW transformation.  As there are $n!$ distinct labelings of an $n$-site lattice, however, such an optimization naively scales combinatorially with system size.  In contrast, optimizing the EJW parameters $\boldsymbol{\theta}$ along with the Hartree-Fock parameters $\boldsymbol{t}$ has the same scaling with system size (albeit with a larger prefactor) as does a single Hartree-Fock calculation with a traditional JW transformation.

\section{Lie Algebraic Similarity Transformation Theory}
\subsection{An Alternative Perspective on Extended Jordan-Wigner Transformation}
One might be uncomfortable with treating Hamiltonian parameters as variational objects.  To show that this is safe in this context, we now wish to provide an alternative perspective on the same basic idea, in which the Hamiltonian is extracted from a standard JW transformation and we have a wave function which depends on both $\boldsymbol{t}$ and a \textit{symmetric} matrix $\boldsymbol{\theta}$. Along the way, we will also see how we can introduce correlations via a slight modification of the transformation.  We have given the previous perspective first as it might perhaps provide greater insight, but a purely wave function--based approach is conceptually simpler and we will use this language from now on.

Consider, then, Lie algebraic similarity transformation theory.  In LAST, we write a wave function
\begin{subequations}
\begin{align}
|\mathrm{LAST}\rangle &= \mathrm{e}^{\alpha_2} |\Psi_0\rangle,
\\
\alpha_2 &= \frac{1}{2} \, \sum_{pq} \alpha_{pq} \, n_p \, n_q
\end{align}
\end{subequations}
where the matrix $\boldsymbol{\alpha}$ is symmetric and $\alpha_{pp} = 0$; $|\Psi_0\rangle$ is a reference wave function which should be expressable as a (short) linear combination of Slater determinants, for reasons which will become apparent presently.  Because using the LAST wave function in a variational manner is computationally forbidding, we follow traditional coupled cluster theory in using the exponential correlator to transform the Hamiltonian, to
\begin{equation}
\bar{H} = \mathrm{e}^{-\alpha_2} \, H \, \mathrm{e}^{\alpha_2},
\end{equation}
and solve a similarity-transformed Schr\"odinger equation.  The similarity transformation can be evaluated by using the commutator expansion, just as in traditional coupled cluster theory:
\begin{equation}
\bar{H} = H + [H,\alpha_2] + \frac{1}{2} \, [[H,\alpha_2],\alpha_2] + \ldots,
\end{equation}

Unlike in standard coupled cluster, the commutator expansion is non-terminating.  Fortunately, it is instead resummable.\cite{Wahlen-Strothman2015}  Because these ideas are somewhat non-standard, we will rederive them here; see Ref. \citenum{Wahlen-Strothman2015} for more details.

\begin{figure}[t]
\includegraphics[width=\columnwidth]{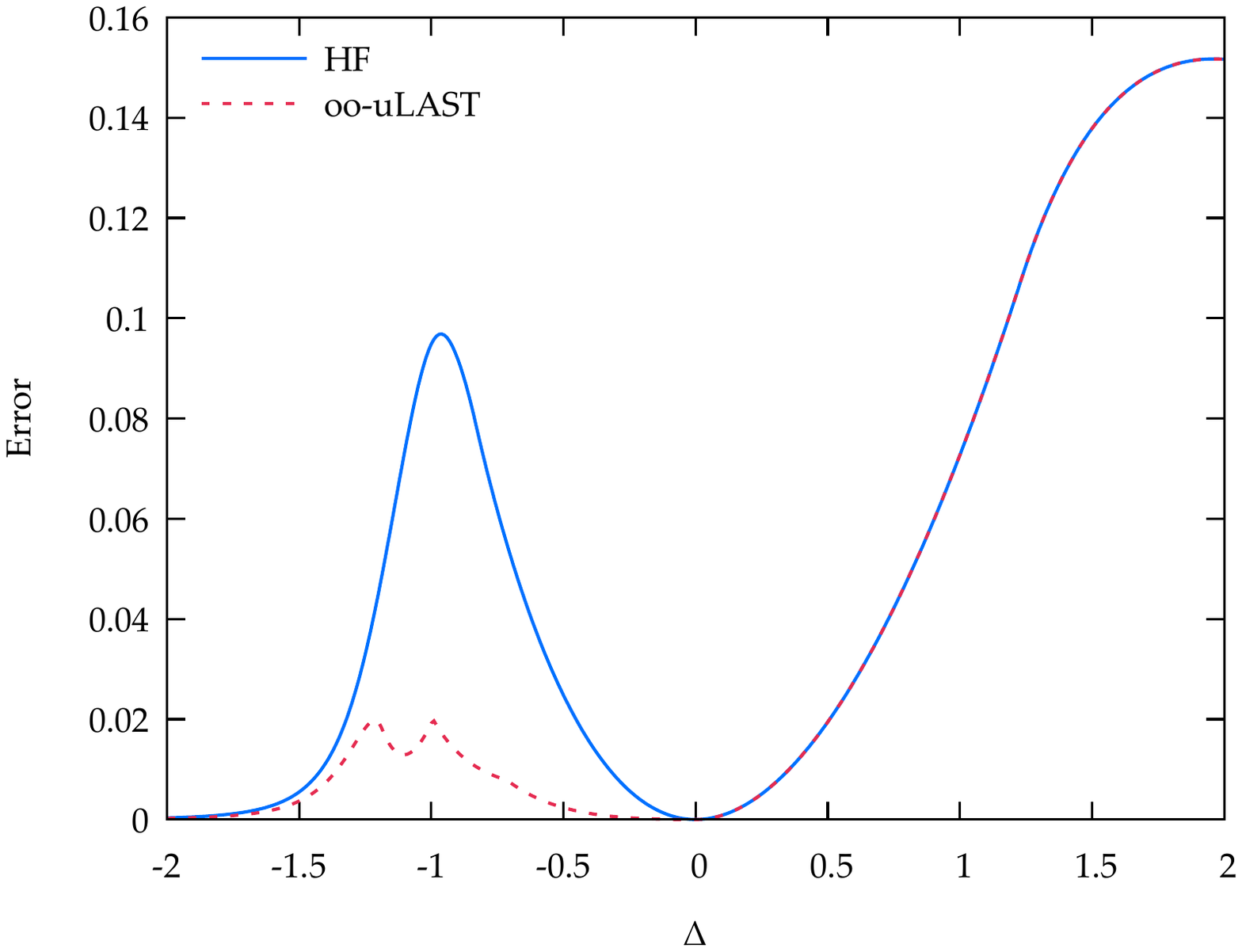}
\caption{HF and oo-uLAST in the JW-transformed 6-site 1D XXZ model with open boundary conditions and $S_z = 0$.  Note that Hartree-Fock and oo-uLAST coincide for $\Delta > 0$.
\label{Fig:XXZOrderingHFvsLAST}}
\end{figure}

Consider, then, the transformation of a single annihilation operator:
\begin{equation}
\bar{c}_p = \mathrm{e}^{-\alpha_2} \, c_p \, \mathrm{e}^{\alpha_2}= \sum \frac{1}{n!} \, \left(c_p \, \alpha_2^n\right)_c
\end{equation}
where the notation $\left(c_p \, \alpha_2^n\right)_c$ means the $n$-fold commutator
\begin{equation}
\left(c_p \, \alpha_2^n\right)_c = \underbrace{[[[[c_p,\alpha_2],\alpha_2],\alpha_2],\ldots]}_{n\textrm{ commutators}}.
\end{equation}
We will use the fact (proven for completeness in Appendix \ref{Sec:AppendixLAST}) that
\begin{subequations}
\begin{align}
\left(c_p \, \alpha_2^n\right)_c &= c_p \, \alpha_{1,p}^n,
\\
\alpha_{1,p} &= \sum_{k} \alpha_{pk} \, n_k.
\end{align}
\end{subequations}
This being the case we see that
\begin{equation}
\bar{c}_p = c_p \, \mathrm{e}^{\alpha_{1,p}}.
\end{equation}
Similarly, one obtains
\begin{equation}
\bar{c}_p^\dagger = c_p^\dagger \, \mathrm{e}^{-\alpha_{1,p}}.
\end{equation}

This means that a generic two-body fermionic Hamiltonian
\begin{equation}
H = \sum h_{pq} \, c_p^\dagger \, c_q + \frac{1}{4} \, \sum h_{pqrs} \, c_p^\dagger \, c_q^\dagger \, c_s \, c_r
\end{equation}
becomes, after similarity transformation,
\begin{align}
\bar{H} &= \sum h_{pq} \, c_p^\dagger \, \mathrm{e}^{-\alpha_{1,p}} \, \mathrm{e}^{\alpha_{1,q}} \, c_q
\\
 &+ \frac{1}{4} \, \sum h_{pqrs} \, c_p^\dagger \, \mathrm{e}^{-\alpha_{1,p}} \, c_q^\dagger \, \mathrm{e}^{-\alpha_{1,q}} \, \mathrm{e}^{\alpha_{1,s}} \, c_s \, \mathrm{e}^{\alpha_{1,r}} \, c_r.
\nonumber
\end{align}
Working with such a Hamiltonian is generally difficult due to the exponential one-body operators, but as in the case of the JW-transformed Hamiltonian, we may readily evaluate matrix elements between Slater determinants since the exponentials of the one-body operators $\mathrm{e}^{\alpha_{1,p}}$ are simply Thouless transformation operators.  It is for this reason that the reference wave function $|\Psi_0\rangle$ must be a single determinant or a short linear combination of determinants.

Now suppose we have a unitary version of LAST (uLAST).  Unitarity means that the symmetric matrix $\boldsymbol{\alpha}$ must be purely imaginary so that the operator $\alpha_2$ is antihermitian.  We thus write $\boldsymbol{\alpha} = \mathrm{i} \, \boldsymbol{\theta}$, where $\boldsymbol{\theta}$ is then real and symmetric, with $\theta_{pp} = 0$, and we choose the reference wave function $|\Psi_0\rangle$ to be a single determinant which we will optimize together with the LAST parameters to get what we will refer to as orbital-optimized uLAST (oo-uLAST).  In this case, the wave function is
\begin{subequations}
\begin{align}
|\mathrm{uLAST}\rangle &= \mathrm{e}^{\mathrm{i} \, \theta_2} \, \mathrm{e}^{T_1} |0\rangle,
\\
\theta_2 &= \frac{1}{2} \, \sum \theta_{pq} \, n_p \, n_q,
\\
T_1 &= \sum t_i^a \, c_a^\dagger \, c_i,
\end{align}
\end{subequations}
for some reference determinant $|0\rangle$.  The expectation value is
\begin{equation}
E_\mathrm{uLAST} = \frac{\langle 0| \mathrm{e}^{T_1^\dagger} \, \mathrm{e}^{-\mathrm{i} \, \theta_2} \, H \, \mathrm{e}^{\mathrm{i} \, \theta_2} \, \mathrm{e}^{T_1} |0\rangle}{\langle 0| \mathrm{e}^{T_1^\dagger} \, \mathrm{e}^{T_1} |0\rangle.}
\end{equation}
This can undoubtedly be minimized, for fixed $H$, with respect to the parameters $\boldsymbol{t}$ and $\boldsymbol{\theta}$.  In practice, we can absorb the parameters $\boldsymbol{\theta}$ into defining a similarity-transformed Hamiltonian $\bar{H}(\boldsymbol{\theta})$ (and in this case, the similarity transformation is also unitary). In the special case that $H$ is the JW transformation of a spin Hamiltonian $H_s$, then $\bar{H}(\boldsymbol{\theta})$ takes precisely the form of the extended JW transformation of $H_s$.

All of this means we have two distinct but equivalent perspectives on the extended JW transformation ideas we have discussed.  On the one hand, we can consider a more complicated transformation from spins to fermions, where the transformation itself depends on a matrix of coefficients $\boldsymbol{\theta}$ satisfying the constraints of Eqn. \ref{Eqn:ThetaConstraints}. We then approximate the ground state of this transformed Hamiltonian as a single determinant, and minimize the energy with respect to both the transformation parameters and the wave function parameters.  On the other hand, we can imagine instead using the standard JW transformation of $H_s$ which we then solve with orbital-optimized unitary LAST.  In either case, the resulting energy is independent of the labeling of the sites in the underlying spin Hamiltonian $H_s$.

\subsection{Lie Algebraic Similarity Transformation as a Correlated Method}
Thus far, we have seen that oo-uLAST provides a different way of thinking about the Hartree-Fock solution of a spin Hamiltonian with an extended JW transformation.  We must note an important but subtle point: the energy minimum of oo-uLAST may not correspond to the Hartree-Fock solution of any JW-transformed Hamiltonian.  That is, unitary LAST eliminates the ordering dependence and may additionally introduce correlations.  From the perspective of extended JW transformations, these correlations occur when the $\boldsymbol{\theta}$ parameters, even starting from the variationally optimal ordering in a standard JW sense, differ from $x$ and $x +\pi$ (we will see whence the $x$ momentarily).  Figure \ref{Fig:XXZOrderingHFvsLAST} demonstrates this for the same 6-site XXZ model we saw previously.

This should perhaps not be too surprising.  After all, LAST was originally introduced as a correlated method. Previous experience, on the other hand, had taught us that unitary LAST is not as accurate as a non-unitary version of the theory (and indeed frequently has no effect without reference reoptimization).  The formal similarity between uLAST on the one hand and JW-transformation on the other does, however, suggest that the non-unitary, similarity-transformed LAST (ST-LAST) as introduced in Ref. \citenum{Wahlen-Strothman2015} is a natural language in which to add correlations atop our other results without dramatically changing the cost.

\begin{figure*}[t]
\includegraphics[width=0.25\textwidth]{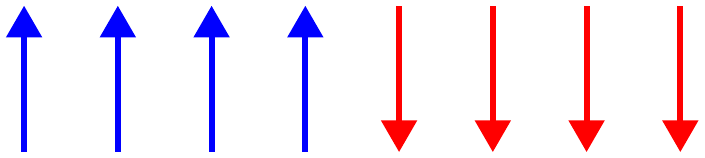}\hfill\includegraphics[width=0.25\textwidth]{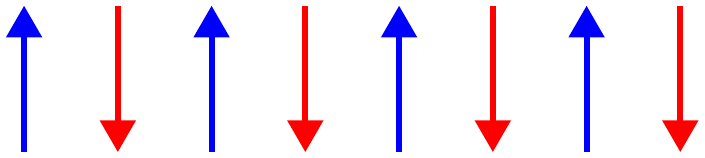}\hfill\includegraphics[width=0.25\textwidth]{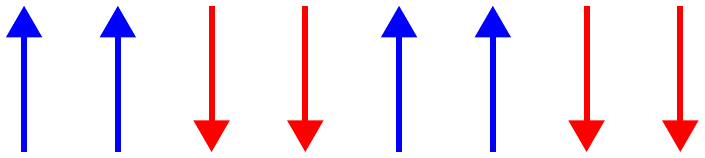}
\\
\vspace{1cm}
\includegraphics[width=0.25\textwidth]{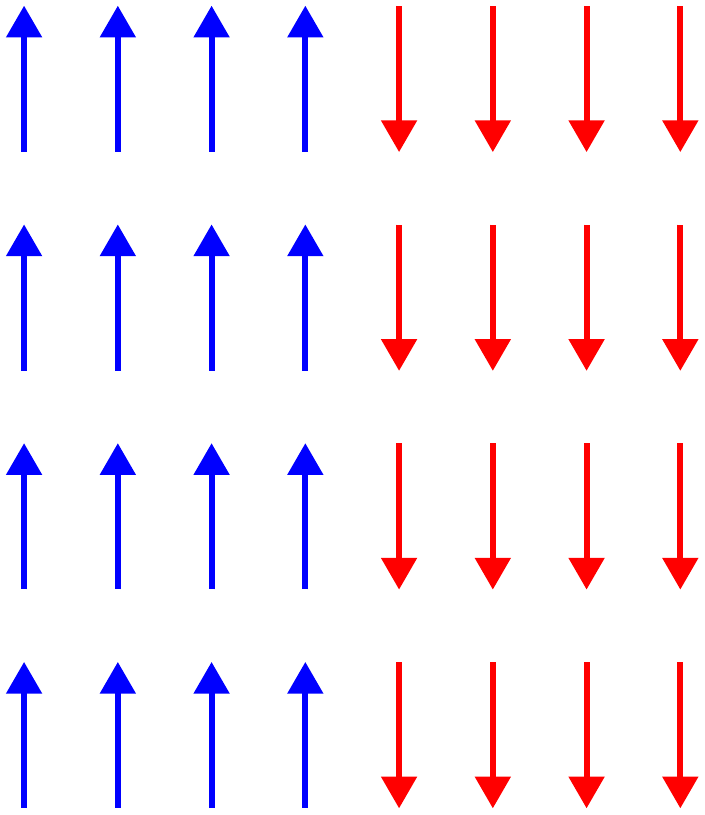}\hfill\includegraphics[width=0.25\textwidth]{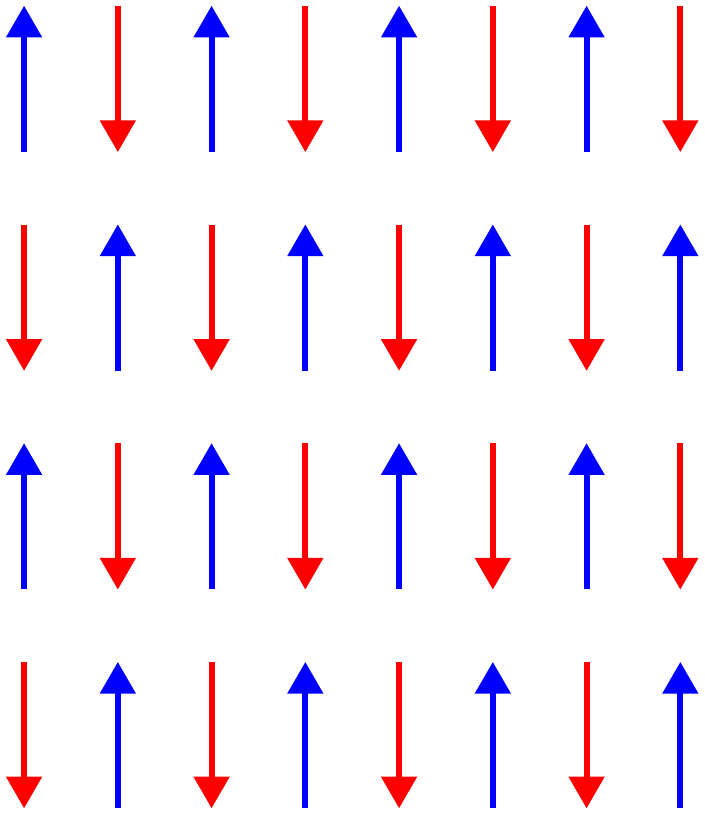}\hfill\includegraphics[width=0.25\textwidth]{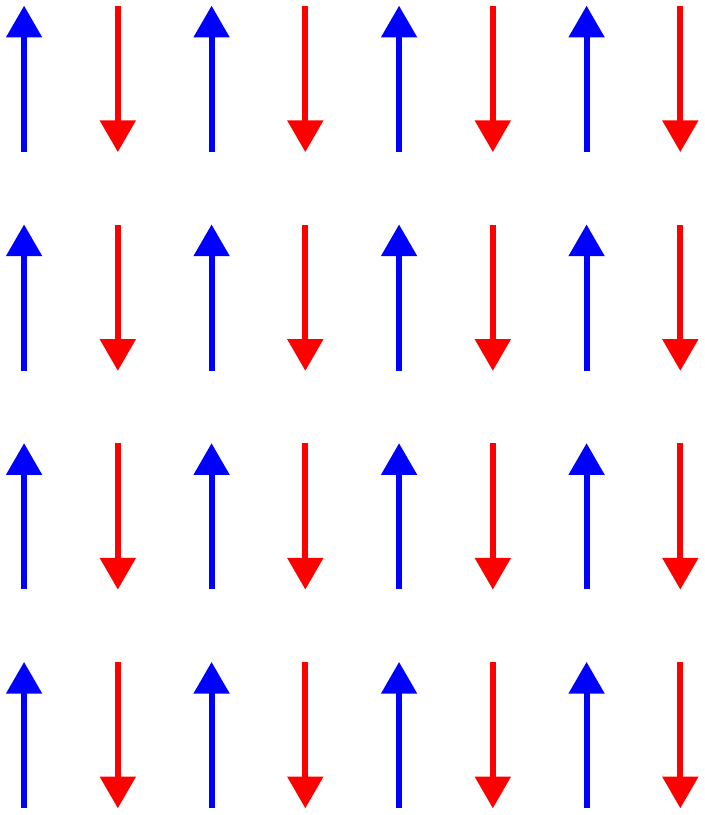}
\caption{Spin arrangments in various 1D and 2D lattices.  
Top row: 1D spin structures.  Top left: Block ferromagnet (for large negative $\Delta$ in the XXZ model).  Top middle: N\'eel structure (large positive $\Delta$ in the XXZ model, small $J_2/J_1$ in the $J_1-J_2$ model).  Top right: Period 2 N\'eel structure (large $J_2/J_1$ in the $J_1-J_2$ model).
Bottom row: 2D spin structures.  Bottom left: Block ferromagnet (for large negative $\Delta$ in the XXZ model).  Bottom middle: N\'eel structure (large positive $\Delta$ in the XXZ model, small $J_2/J_1$ in the $J_1-J_2$ model).  Bottom right: Striped structure (large $J_2/J_1$ in the $J_1-J_2$ model).
\label{Fig:SpinArrangements}}
\end{figure*}

To add correlations with ST-LAST, we begin by variationally minimizing the oo-uLAST energy of the JW-transformed Hamiltonian to obtain the parameters $\boldsymbol{\theta}$ and $\boldsymbol{t}$.  Once we have done so, we incorporate additional correlations by solving the standard ST-LAST equations:\cite{Wahlen-Strothman2015}
\begin{subequations}
\begin{align}
E &= \frac{\langle \Phi| \tilde{H} |\Phi\rangle}{\langle \Phi|\Phi\rangle},
\\
0 &= \langle \Phi| n_p \, n_q \, (\tilde{H} - E) |\Phi\rangle,
\label{Eqn:STLASTAmplitude}
\end{align}
\end{subequations}
where
\begin{subequations}
\begin{align}
|\Phi\rangle &= \mathrm{e}^{T_1} |0\rangle,
\\
\tilde{H} &= \mathrm{e}^{-\alpha_2 - \mathrm{i} \, \theta_2} \, H \, \mathrm{e}^{\alpha_2 + \mathrm{i} \, \theta_2},
\end{align}
\label{Eqn:STLastDefs}
\end{subequations}
and where we set $T_1$ and $\theta_2$ to their oo-uLAST values and obtain the coefficients in $\alpha_2$ from the amplitude equation of Eqn. \ref{Eqn:STLASTAmplitude}.  In practical implementation, we can absorb both $\alpha_2$ and $\mathrm{i} \, \theta_2$ into a non-unitary extended JW string.

Finally, we must note that there is a redundancy in the LAST parameterization of the wave function (and thus also in the extended JW transformation).  If we shift all $\alpha_{pq}$ for $p \ne q$ by a constant $\delta$, then we shift $\alpha_2$ in a trivial way:
\begin{align}
\alpha_2 = \frac{1}{2} \, \sum_{p \ne q} \alpha_{pq} \, n_p \, n_q
 &\to \frac{1}{2} \, \sum_{p \ne q} \left(\alpha_{pq} + \delta\right) \, n_p \, n_q
\\
 &= \alpha_2 + \frac{1}{2} \, \delta \, \sum_{p \ne q} n_p \, n_q
\nonumber
\\
 &= \alpha_2 + \frac{1}{2} \, \delta \, \left(N^2 - N\right)
\nonumber
\end{align}
where $N$ is the fermionic total number operator.  Acting on number eigenstates, shifting all the LAST parameters by a constant just multiplies the LAST wave function by a constant and does not change the similarity-transformed Hamiltonian at all.  It is this constant that gives rise to the factor of $x$ we mentioned at the beginning of this subsection.

\section{Results}
We have seen that oo-uLAST yields results which are invariant to the labeling of sites in a lattice and may, in addition, provide correlation.  Here, we wish to explore the practical consequences of treating spin Hamiltonians with oo-uLAST.  We also wish to investigate the degree to which we can supplement oo-uLAST with non-unitary similarity-transformed LAST, as previously described.  To this end, we will consider two spin Hamiltonians: the XXZ model given in Eqn. \ref{Eqn:DefHXXZ} and the $J_1-J_2$ model 
\begin{equation}
H_{J_1-J_2} = J_1 \, \sum_{\langle pq \rangle} \vec{S}_p \cdot \vec{S}_q + J_2 \, \sum_{\langle\langle pq \rangle \rangle} \vec{S}_p \cdot \vec{S}_q
\label{Eqn:DefHJ1J2}
\end{equation}
in which nearest neighbor sites (denoted by $\langle pq \rangle$) have an isotropic Heisenberg interaction with coefficient $J_1$, and next-nearest-neighbor sites (denoted by $\langle \langle pq \rangle\rangle$) interact with strength $J_2$.  We will take $J_1 = 1$.  In both Hamiltonians, each site is spin 1/2.

\begin{figure*}[t]
\includegraphics[width=0.45\textwidth]{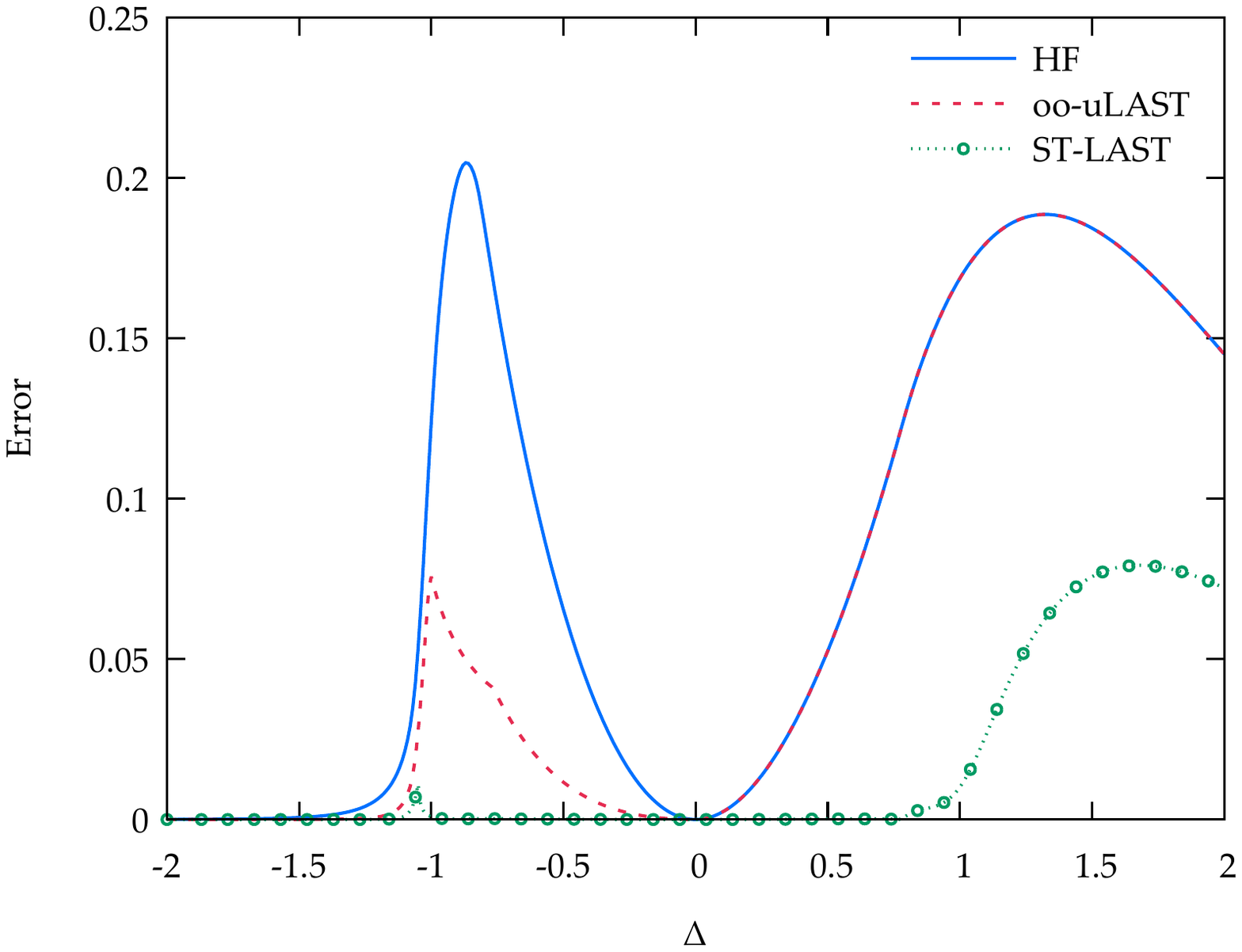}
\hfill
\includegraphics[width=0.45\textwidth]{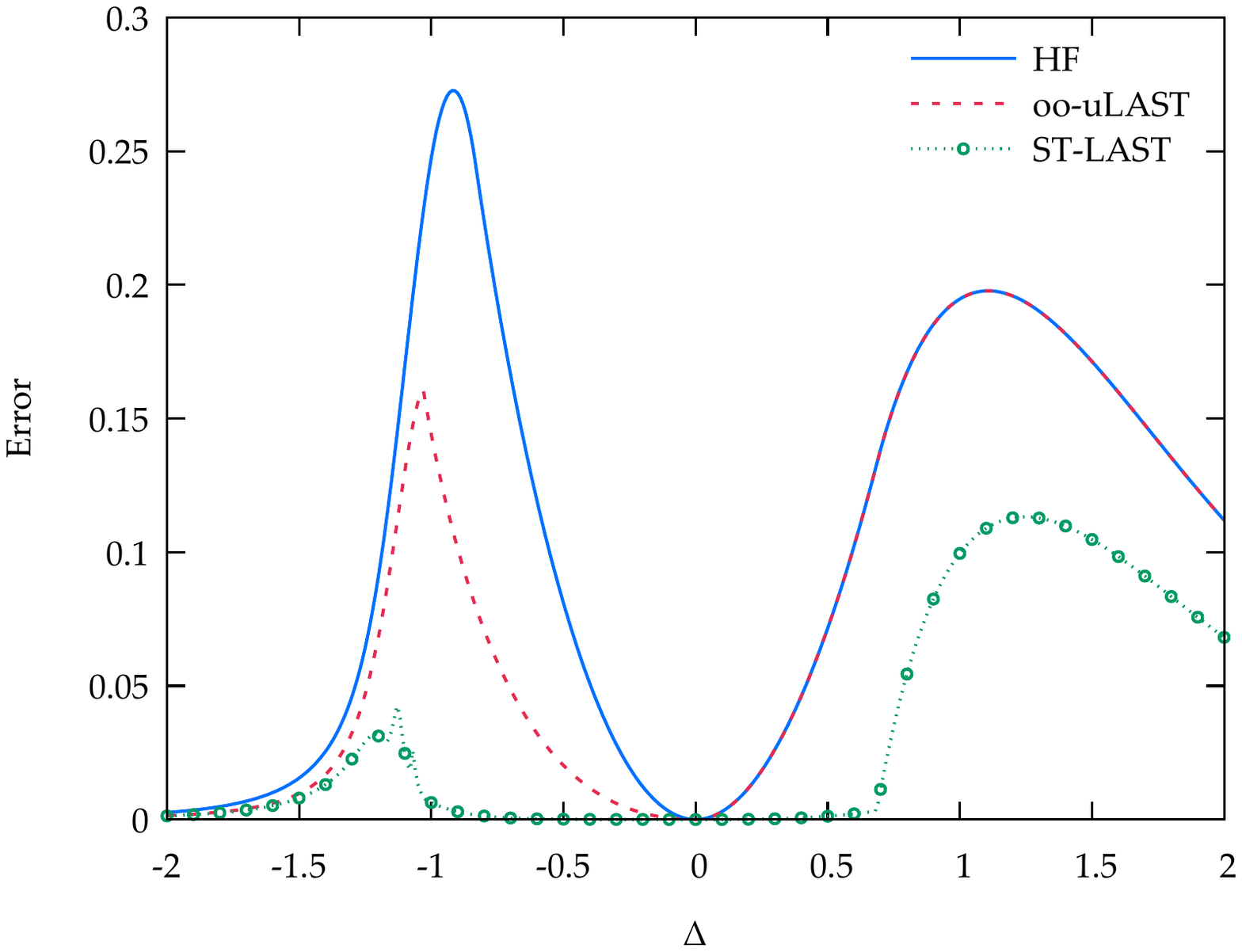}
\\
\caption{Energy errors in the 12-site 1D XXZ Hamiltonian with $S_z = 0$.  Left panel: Open boundary conditions.  Right panel: Periodic boundary conditions.  Note that for $\Delta \ge 0$, oo-LAST and HF coincide.
\label{Fig:1DXXZ}}
\end{figure*}

\begin{figure*}
\includegraphics[width=0.45\textwidth]{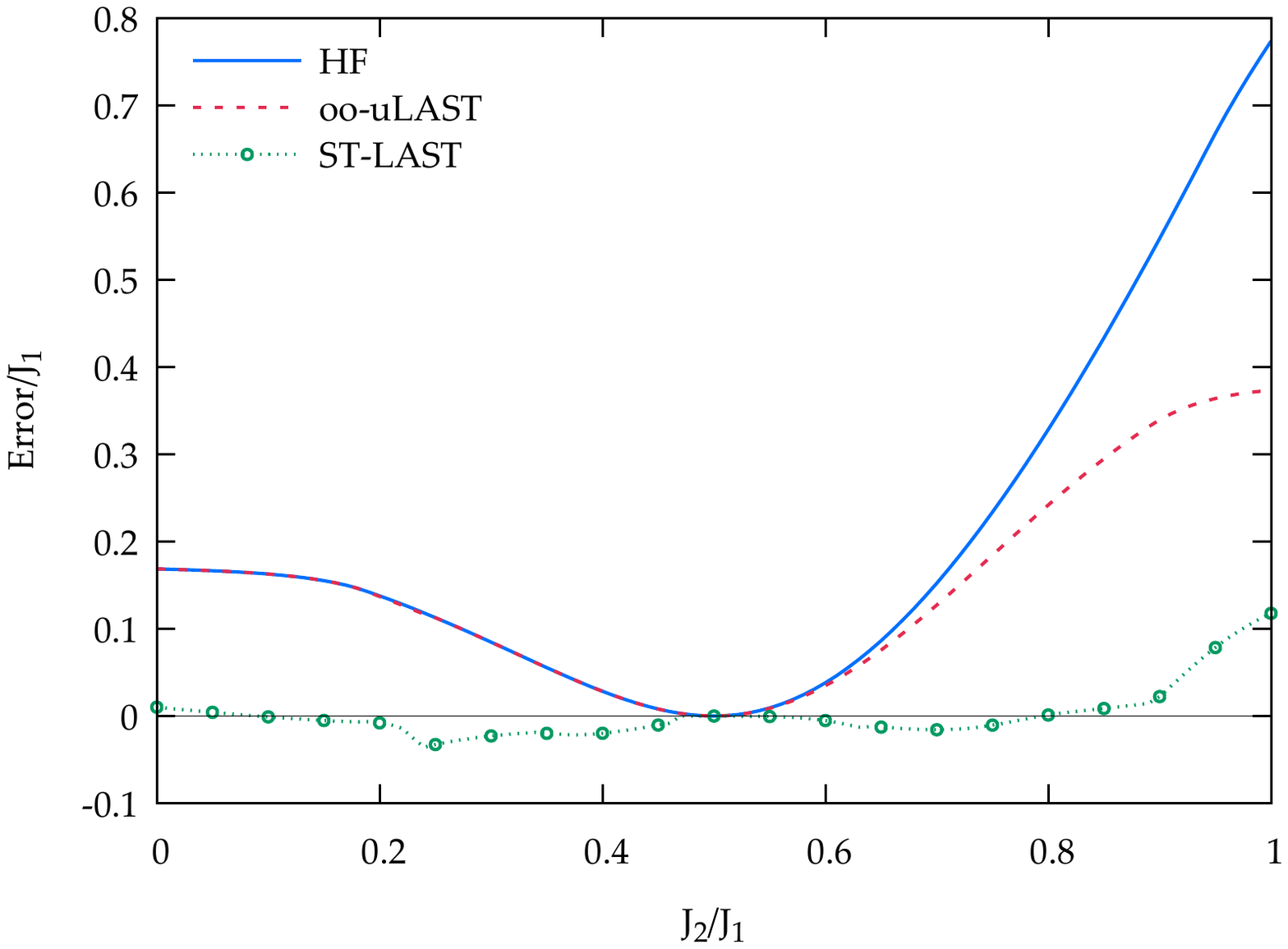}
\hfill
\includegraphics[width=0.45\textwidth]{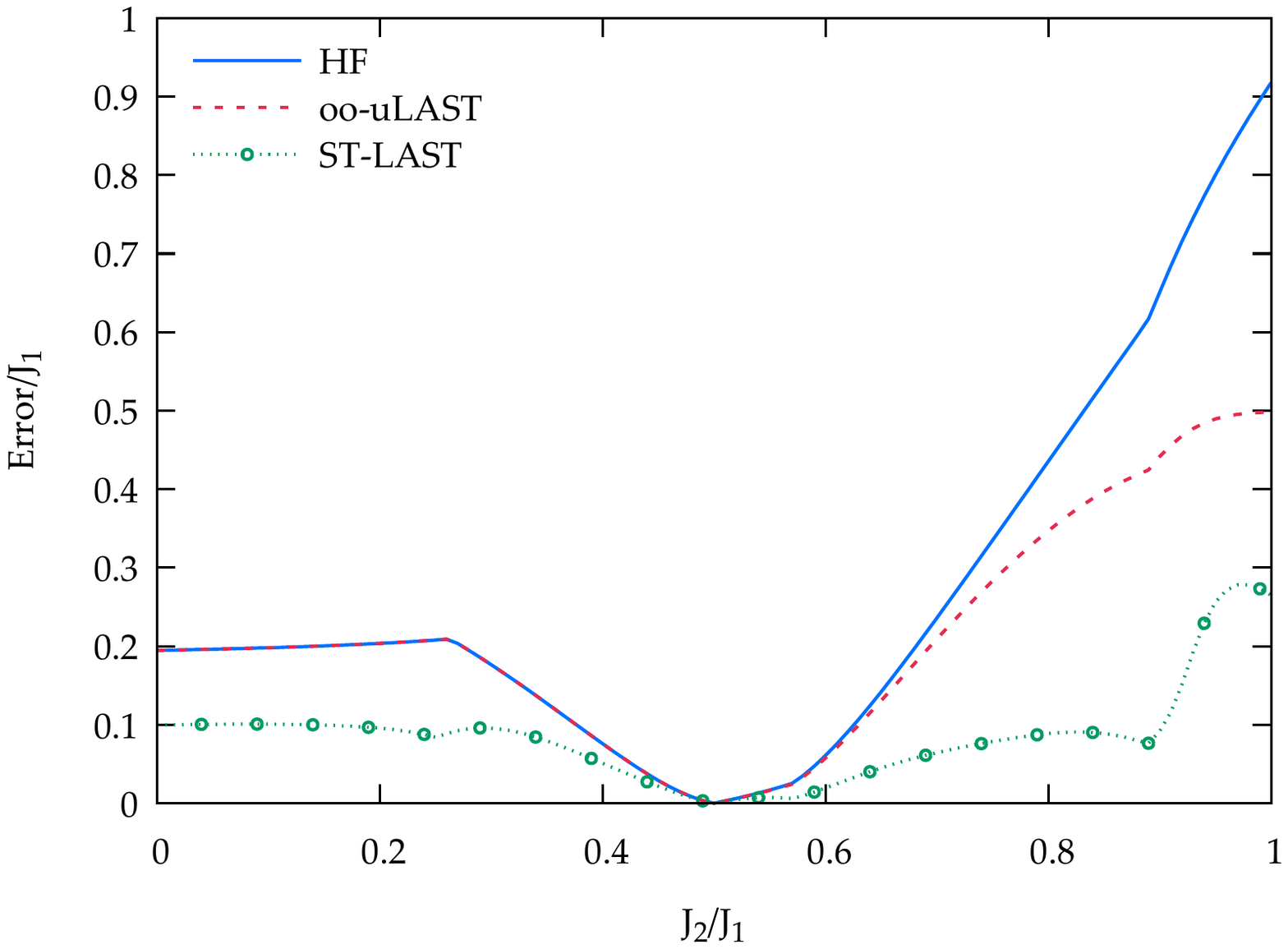}
\caption{Energy errors in the 12-site 1D $J_1-J_2$ Hamiltonian with $S_z = 0$.  Left panel: Open boundary conditions.  Right panel: Periodic boundary conditions.  Note that for $J_2/J_1 \le 1/2$, oo-LAST and HF coincide.
\label{Fig:1DJ1J2}}
\end{figure*}

Before we discuss our detailed numerical results, we must say a few words about the model Hamiltonians we will examine.  The XXZ model has, in the thermodynamic limit, three phases with $\langle S_z \rangle = 0$.  For $\Delta < -1$, the ground state is a block ferromagnet, in which each site in one half of the lattice is $\uparrow$-spin and each site in the other half is $\downarrow$-spin.  For $\Delta > 1$, the ground state is instead a N\'eel antiferromagnet, while for $|\Delta| < 1$ the ground state is an XY phase in which each site has, on average, $\langle S_z \rangle = 0$ with the magnetism oriented in the $x-y$ plane.\cite{Farnell2004}  Similarly, the $J_1-J_2$ model has (in two dimensions) a N\'eel antiferromagnetic ground state for $J_2 \lesssim 0.4$, and striped antiferromagnet for $J_2 \gtrsim 0.6$.  For intermediate values of $J_2$, the nature of the ground state is still under a certain amount of discussion.\cite{Darradi2008,Gong2014,Richter2015} In one dimension, the $J_1-J_2$ model with $J_1 > 0$ has a N\'eel antiferromagnetic ground state for $J_2 \lesssim 1/2$ and a frustrated phase for $J_2 \gtrsim 1/2$.\cite{Bishop1998}

We study finite systems in this work, for which there are of course no phase transitions.  Nonetheless, one still observes multiple kinds of magnetic orderings which in the exact wave function evolve smoothly from one to another as we adjust parameter values ($\Delta$ or $J_2$ for the XXZ or $J_1-J_2$ model, respectively).  With Hartree-Fock and oo-uLAST, we generally see solutions which cross one another instead.  Similarity-transformed LAST solutions also cross one another, though not necessarily at the same parameter values at which the oo-LAST reference states cross.  Note that because we do not impose any kind of lattice symmetry, the frustrated phase in the 1D $J_1-J_2$ model becomes a kind of ``period 2'' N\'eel arrangement as $J_2$ dominates over $J_1$.  Figure \ref{Fig:SpinArrangements} show the various magnetic structures in our various model Hamiltonians.

\begin{figure*}[t]
\includegraphics[width=0.45\textwidth]{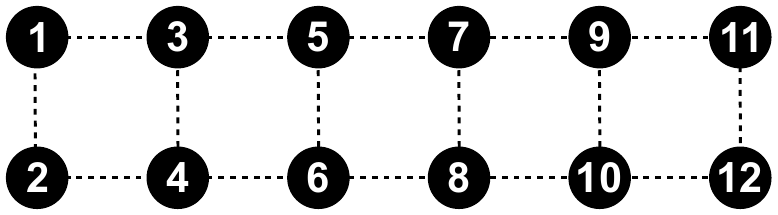}
\hfill
\includegraphics[width=0.45\textwidth]{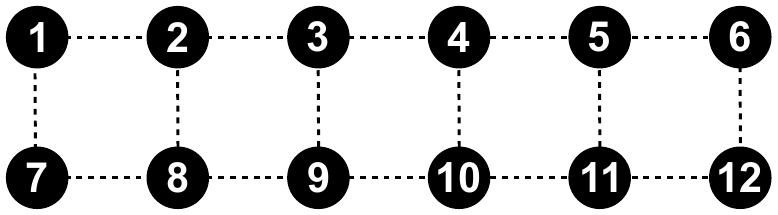}
\caption{Site labeling in quasi-1D systems.  Left panel: Labeling for a $2 \times 6$ lattice. Right panel: Labeling for a $6 \times 2$ lattice.
\label{Fig:LadderLabeling}}
\end{figure*}

We should also briefly touch on the way in which we solve the equations for the wave function amplitudes $t_i^a$, $\theta_{pq}$, and $\alpha_{pq}$.  Although the most computationally efficient implementation uses a non-orthogonal version of Wick's theorem\cite{Balian1969,Chen2023} to evaluate matrix elements of the Hamiltonian between two different Slater determinants, our testing implementation here uses a full configuration interaction code instead.  This facilitates implementation and testing but limits us to 12-16 sites, even though the formal scaling of our energy evaluation for Hamiltonians of the form given in Eqn. \ref{Eqn:DefHSpin} is $\mathcal{O}(N^5)$, where $N$ is the number of sites.  This scaling arises because evaluating the energy for a generic two-body spin Hamiltonian requires $\mathcal{O}(N^2)$ distinct matrix elements, each of which may be evaluated in $\mathcal{O}(N^3)$ time.

Given the energy and the analytic gradient, we begin by choosing a single lattice-basis determinant from those with the lowest energy at some extreme value of the parameter (e.g. $\Delta = \pm 2$ or $J_2$ = 0 or 1) and initialize the $t_i^a$ amplitudes of Hartree-Fock to zero; we then minimize the energy using the conjugate gradient algorithm, and use the converged solution from one parameter value as the initial guess for the next.  We take a similar approach for oo-uLAST, in which we initialize $t_i^a$ to zero and $\theta_{pq}$ to random numbers.  For ST-LAST, we take an approach inspired by traditional coupled cluster theory. At each parameter value, we initialize $\alpha_{pq} = 0$ and compute the diagonal of the Jacobian:
\begin{equation}
J_{pq,pq} = \left.\frac{\partial R_{pq}}{\partial \alpha_{pq}}\right\rvert_{\alpha_2 = 0}
\end{equation}
where the residual is
\begin{equation}
R_{pq} = \langle \Phi| n_p \, n_q \, (\tilde{H} - E) |\Phi\rangle
\end{equation}
in terms of the state $|\Phi\rangle$ and transformed Hamiltonian $\tilde{H}$ defined in Eqn. \ref{Eqn:STLastDefs}.  We use this diagonal approximation in a quasi-Newton scheme, iteratively solving
\begin{subequations}
\begin{align}
\alpha_{pq} &\to \alpha_{pq} + \delta \alpha_{pq},
\\
0 &= R_{pq} + J_{pq,pq} \, \delta \alpha_{pq},
\end{align}
\end{subequations}
and accelerating convergence using the DIIS algorithm.\cite{Scuseria1986}  Once the residual is sufficiently small, we switch to a standard Newton-Raphson approach.  To handle the singularities implied by the invariance of LAST to shifting $\alpha_{pq}$ by a constant, we set $\langle \alpha_2 \rangle = 0$ at each iteration and regularize the linear equations
\begin{equation}
\boldsymbol{J}^{-1} \to \left(\boldsymbol{J} + \omega \, \boldsymbol{1}\right)^{-1} \, \boldsymbol{J} \, \left(\boldsymbol{J} + \omega \, \boldsymbol{1}\right)^{-1}
\end{equation}
for $\omega$ some small positive constant (in this case, $10^{-8}$).

Finally, we must specify an ordering for the sites.  We order sites sequentially along the $x$ direction, then the $y$ direction, and do not use a serpentine labeling of the lattice.  Thus, for an $n_x \times n_y$ lattice, the label $i_{x,y}$ of site $(x,y)$ is
\begin{equation}
i_{x,y} = x + (y-1) \, n_x.
\label{Eqn:DefIndex}
\end{equation}

\subsection{One-Dimensional Hamiltonians}
Let us begin with one-dimensional (1D) results.  Figure \ref{Fig:1DXXZ} shows results for the 12-site XXZ model in open boundary conditions (OBC) and periodic boundary conditions (PBC), while Fig. \ref{Fig:1DJ1J2} does the same for the 12-site $J_1-J_2$ model.  

In the XXZ model, HF and oo-uLAST coincide for $\Delta \ge 0$.  In other words, in this region the oo-uLAST restores the permutational invariance of the transformation, but does not add any correlation.  On the other hand, HF and oo-uLAST differ significantly for $\Delta < 0$ and it would appear that oo-uLAST is therefore providing at least some degree of correlation.  These results are exactly analogous to what we see for the 6-site XXZ model in Fig. \ref{Fig:XXZOrderingHFvsLAST}.  While for 12 sites we cannot exclude the possiblity that some ordering of lattices sites exists for which HF and oo-uLAST agree for $\Delta < 0$ (because there are, with open boundary conditions, $1/2 \times 12! \sim 2.4 \times 10^8$ permutations to check), we have checked that in 6 sites and periodic boundary conditions, oo-uLAST does not match the HF result for any lattice ordering.  

We have found three distinct oo-uLAST solutions in OBC: one for $\Delta \gtrsim -0.75$, a second for $\Delta \lesssim -1$, and a third for $\Delta$ between $-0.75$ and $-1$.  In PBC, we find only two solutions.

Turning to the ST-LAST results, we see that ST-LAST greatly improves results except in what is roughly the N\'eel regime ($\Delta \gtrsim 1$), where its effects are markedly smaller.  There appears to be roughly three distinct solutions, although we have had considerable difficulty in resolving where exactly the solutions cross.  Of course HF is exact at $\Delta = 0$, as we discussed earlier, and thus so too are oo-uLAST and ST-LAST.  Note that as we approach large $|\Delta|$, the XXZ Hamiltonian of Eqn. \ref{Eqn:DefHXXZ} is dominated by the $\Delta \, S^z \, S^z$ term or, upon JW-transformation, by a term $\Delta \, \bar{n} \, \bar{n}$ (see Eqn. \ref{Def:HXXZfermionic}).  Mean-field states, whether in terms of spins or fermions, are eigenstates of this term, so in the limit of very large $|\Delta|$, JW-HF is again exact.

Results for the 1D $J_1-J_2$ model are qualitatively similar.  Again, there is a point ($J_2 = 1/2$) at which Hartree-Fock is already exact.  At this point, the system is fully dimerized.\cite{Majumdar1969} For $J_2 < 1/2$, HF and oo-uLAST are identical, while for $J_2 > 1/2$ they differ.  Again, ST-LAST offers significant improvement upon oo-uLAST, although the improvement is smaller than in the XXZ model.  Finally, we note that we observe several solutions, particularly in the periodic case.  This plethora of stationary points of the oo-uLAST energy appears to be the method's principle drawback.

\begin{figure*}[t]
\includegraphics[width=0.45\textwidth]{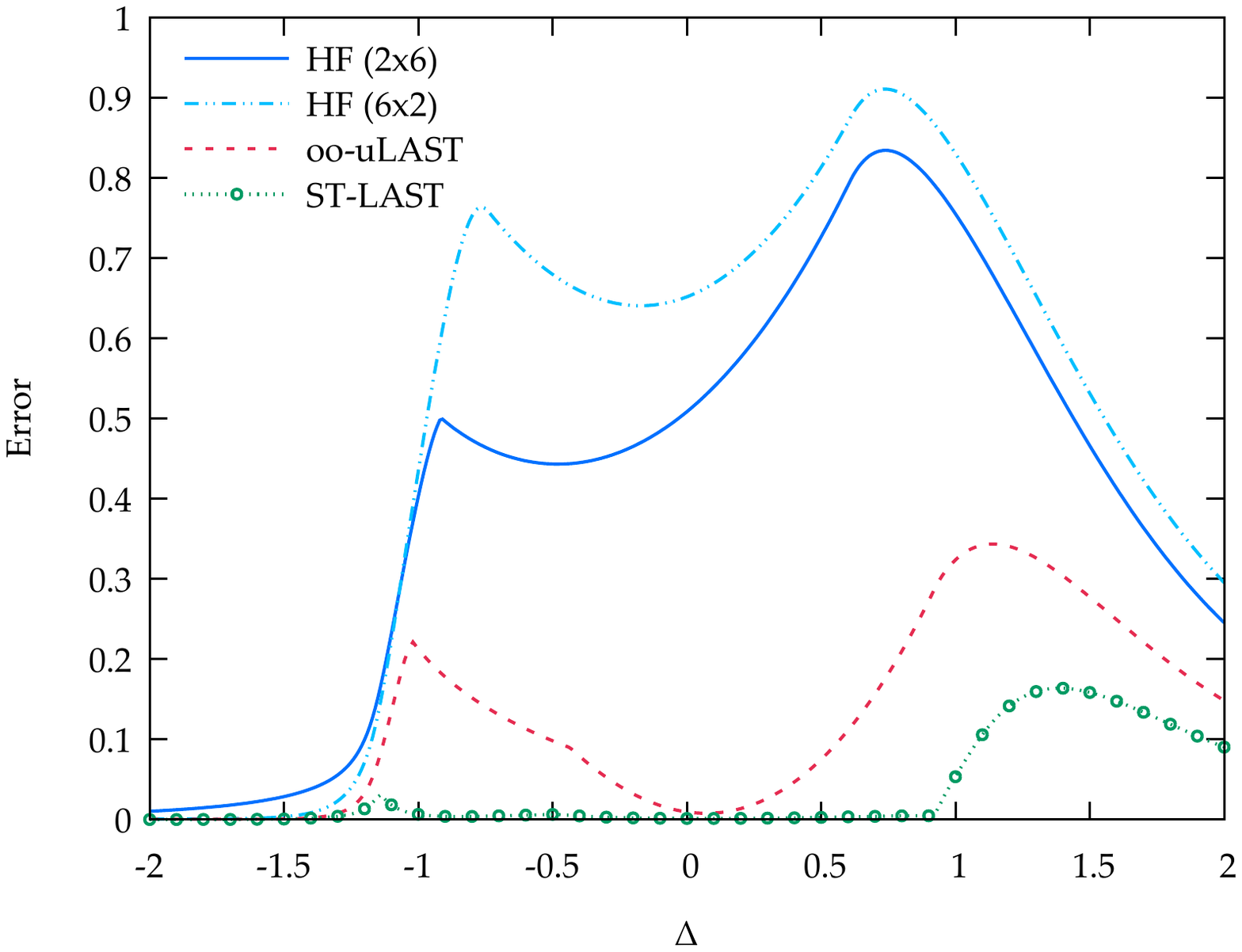}
\hfill
\includegraphics[width=0.45\textwidth]{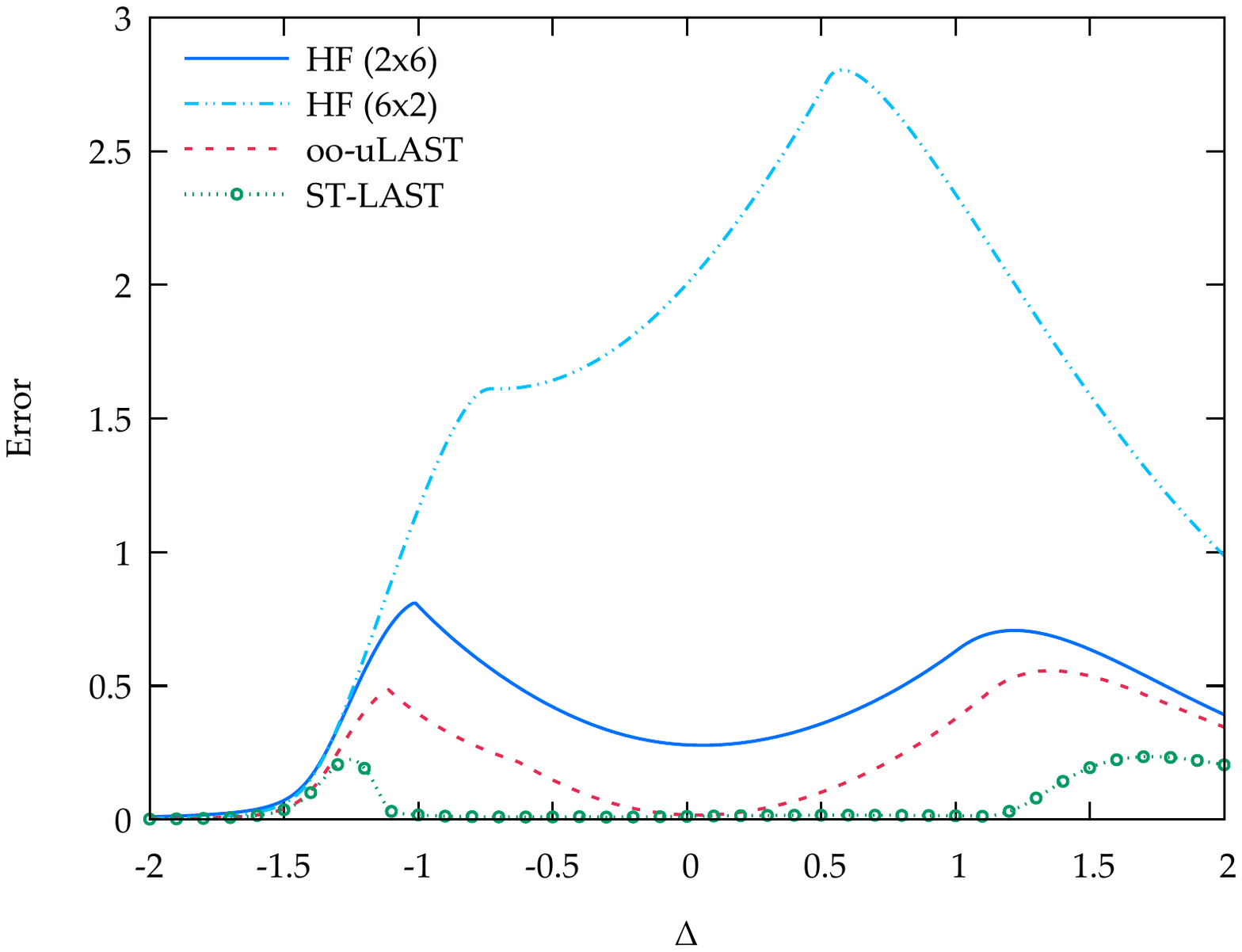}
\caption{Energy errors in XXZ ladder models with $S_z = 0$.  Left panel: Open boundary conditions.  Right panel: Periodic boundary conditions.
\label{Fig:XXZLadder}}
\end{figure*}

\begin{figure*}[t]
\includegraphics[width=0.45\textwidth]{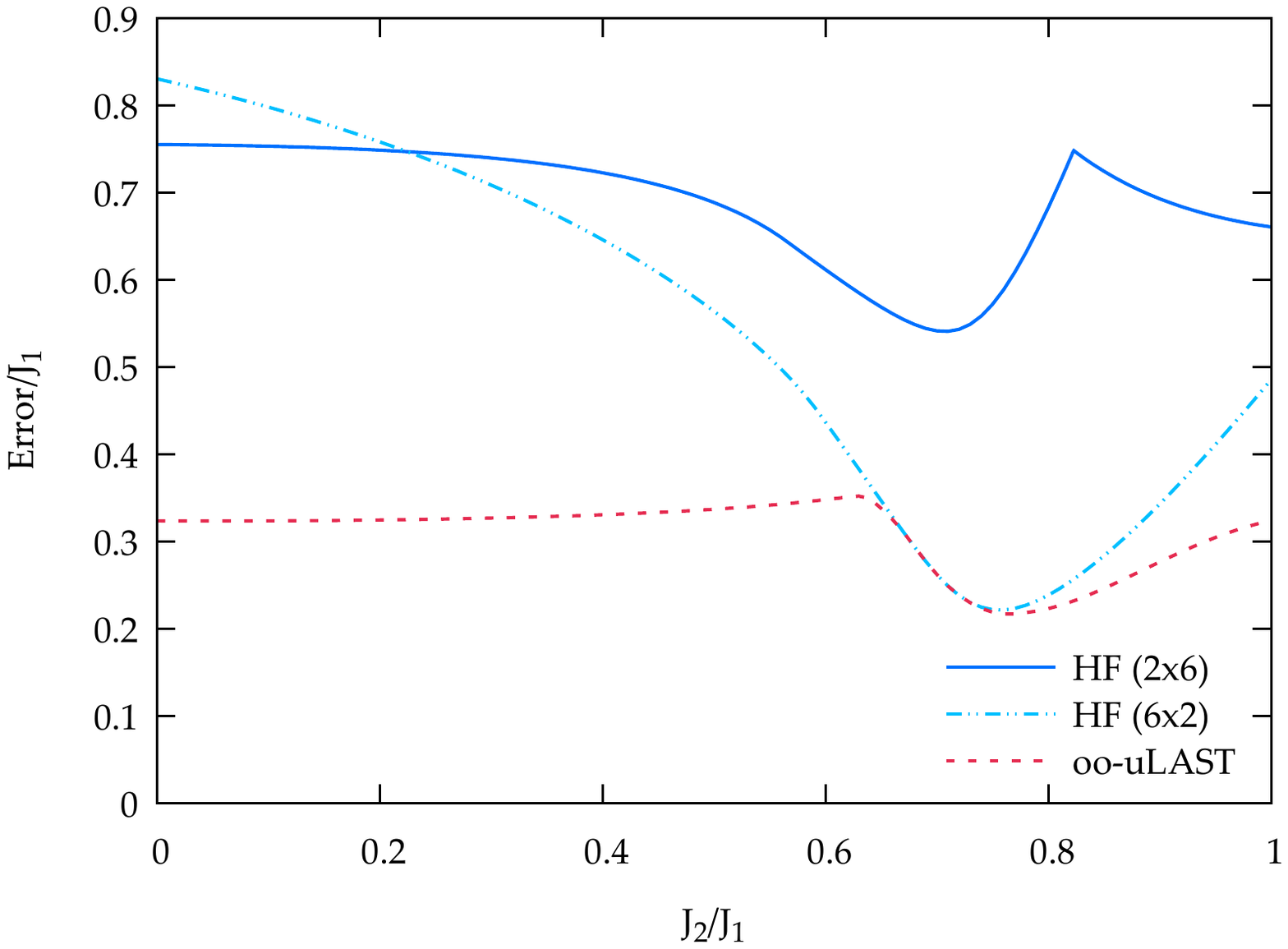}
\hfill
\includegraphics[width=0.45\textwidth]{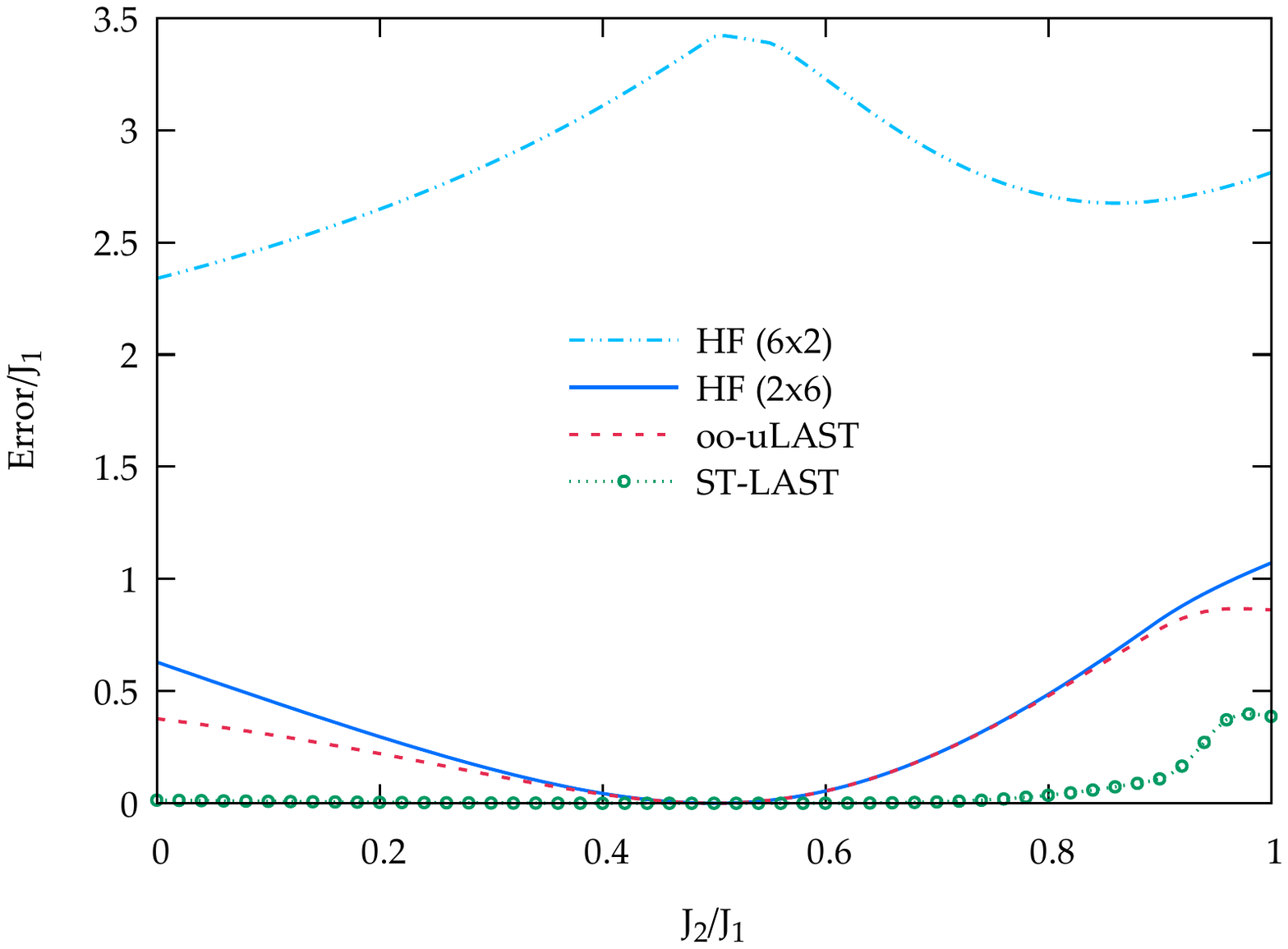}
\caption{Energy errors in $J_1-J_2$ ladder models with $S_z = 0$.  Left panel: Open boundary conditions.  Right panel: Periodic boundary conditions.  We have been unable to converge ST-LAST for the $J_1-J_2$ ladder with open boundary conditions.  Hartree-Fock is exact for the periodic ladder at $J_2 = 1/2 \, J_1$.
\label{Fig:J1J2Ladder}}
\end{figure*}

\subsection{Quasi-One-Dimensional Hamiltonians}
In one dimension, there is an apparently natural order for the lattice sites (we can just label them sequentially) but in two dimensions, this is not the case.  To see this in action, we consider quasi-1D ``ladder'' systems in which the lattice is $2 \times n$ or $n \times 2$ and in which the JW strings with the labeling we have identified in Eqn. \ref{Eqn:DefIndex} differ; we depict these two different labeling schemes in Fig. \ref{Fig:LadderLabeling} for clarity.

We do not wish to belabor the point, but as Figs. \ref{Fig:XXZLadder} and \ref{Fig:J1J2Ladder} make clear, we obtain very different Hartree-Fock results, depending on the labeling.  This ambiguity is completely resolved in oo-uLAST, which additionally appears to provide a substantial degree of correlation in the XXZ model.  The inclusion of a non-unitary similarity-transformed LAST provides quantitatively accurate total energies in the XXZ ladder Hamiltonian, at least for the region $|\Delta| \lesssim 1$ which can be rather difficult to capture in the language of spins.  For the $J_1-J_2$ model, we see the same basic features, although note that we have been unable to systematically converge the ST-LAST equations for the $J_1-J_2$ ladder Hamiltonians with open boundary conditions.  We note also that in the $J_1-J_2$ ladder with periodic boundary conditions, HF is exact at $J_2 = 1/2$ (for some choice of lattice labeling) as are, of course, oo-uLAST and ST-LAST.  As in the 1D case, JW-HF becomes exact in the large $|\Delta|$ limit.

\begin{figure}[b]
\includegraphics[width=0.45\textwidth]{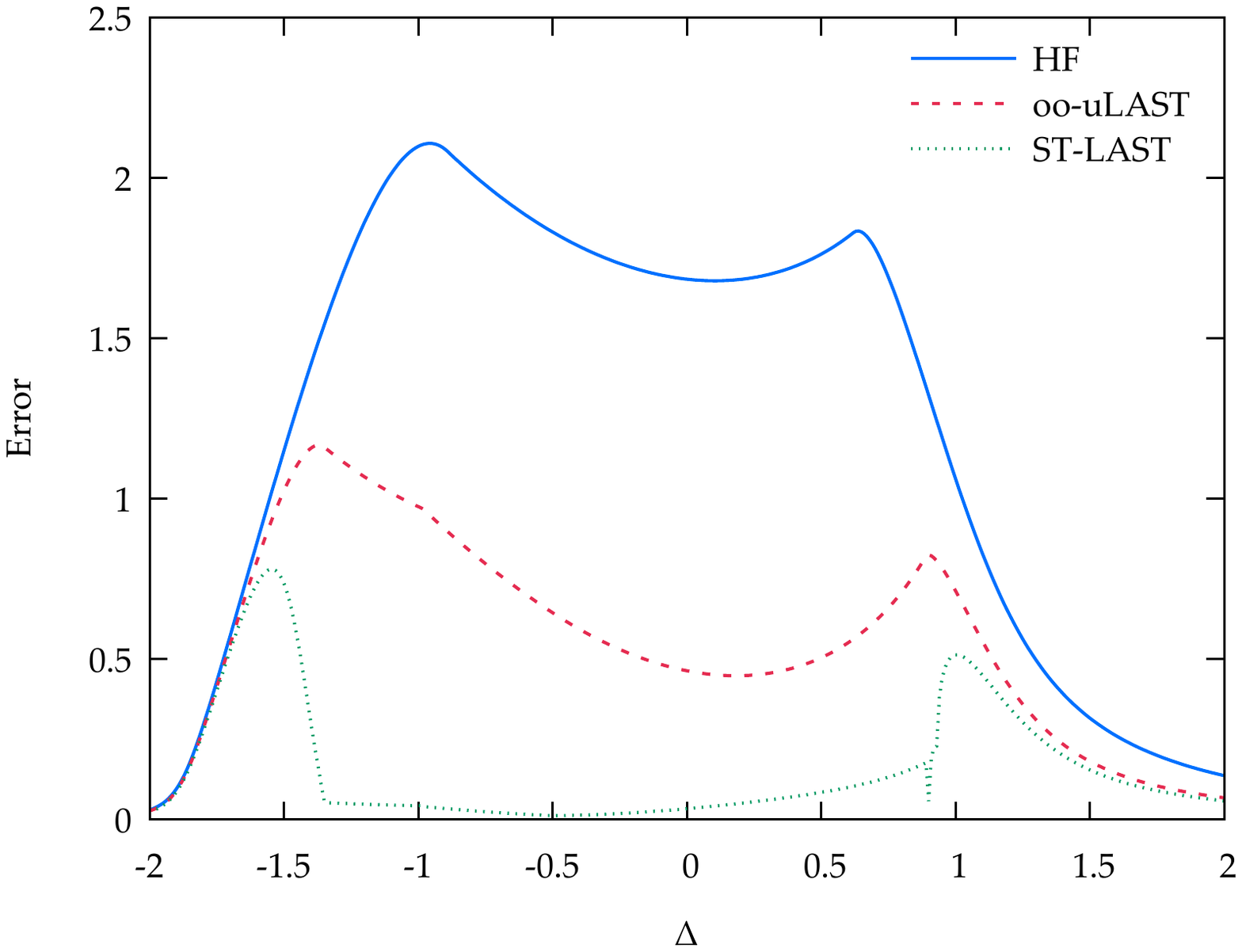}
\hfill
\includegraphics[width=0.45\textwidth]{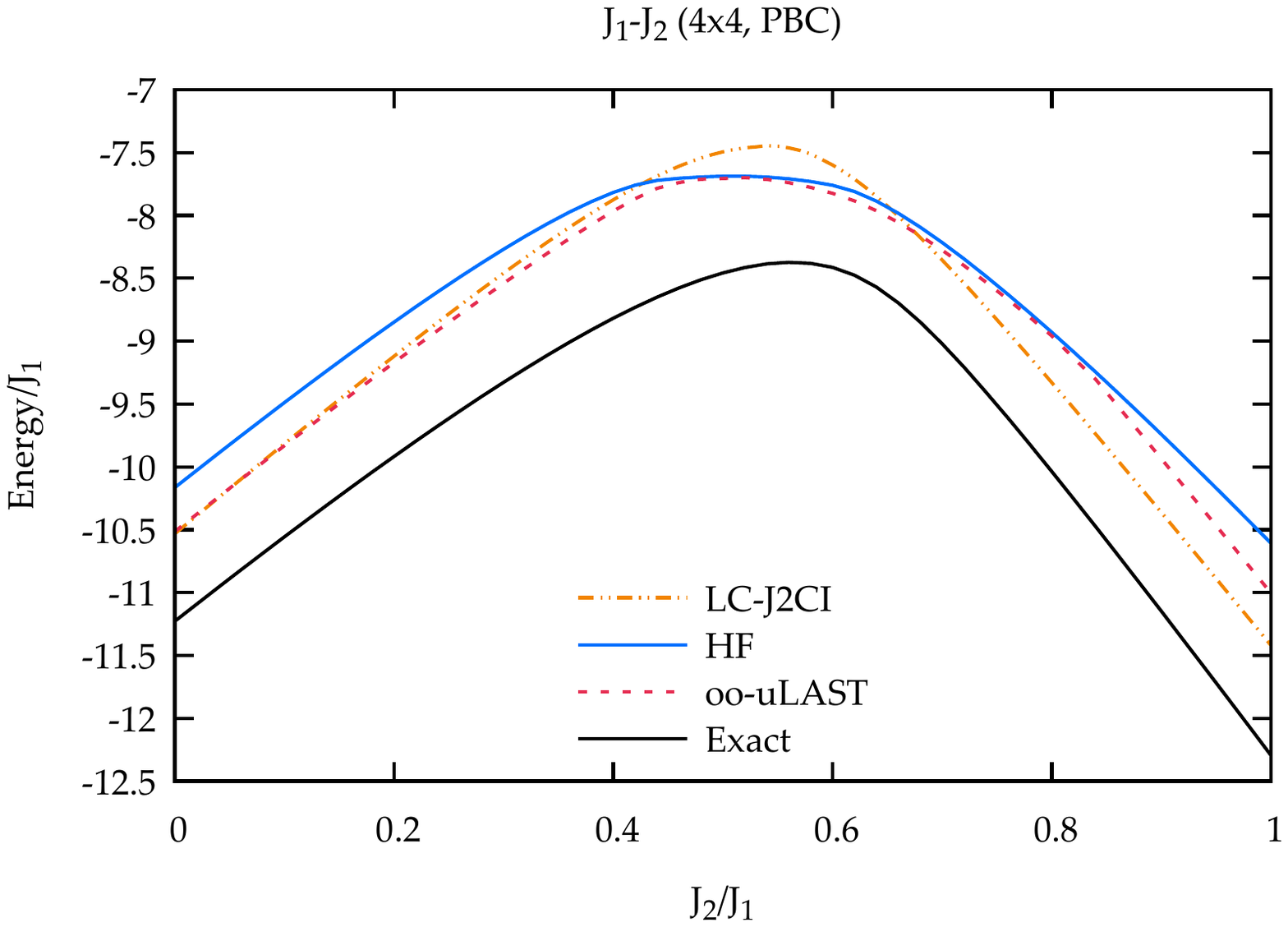}
\caption{Left panel: Energy errors in the 4$\times$4  periodic XXZ Hamiltonian with $S_z = 0$.  Right panel: Total energies in the 4$\times$4 periodic $J_1-J_2$ Hamiltonian with $S_z = 0$.
\label{Fig:2D}}
\end{figure}

\subsection{Two-Dimensional Hamiltonians}
Finally, Fig. \ref{Fig:2D} shows results for the $4 \times 4$ XXZ and $J_1-J_2$ Hamiltonians with periodic boundary conditions.  Clearly, our HF results are worse in $2\times n$ systems than they are in 1D, and worse yet in genuinely 2D systems, but oo-uLAST affords significantly improved performance.  Once we include ST-LAST for correlation, the energetic accuracy is genuinely quite high, although we note that there are many solutions which differ only slightly in energy from one another and whose presence complicates the generation of a complete curve for energy as a function of $\Delta$.  We have not included ST-LAST in the $J_1-J_2$ model, as we have not been able to converge to a real energy at every point, and rather than showing energy errors we plot total energies.  We have also included a curve we call ``LC-J2CI,'' which is a method which works directly in the language of spins.\cite{Liu2023}  This method is a linear combination of seven configuration interaction doubles based upon the spin version of the antisymmetrized geminal power wave function.  In the language of conventional electronic structure methods, one might think of it as a linear combination of multireference configuration interaction states.  It is remarkable that even a simple Hartree-Fock solution of the JW-transformed Hamiltonian is of roughly comparable quality.

\begin{figure*}
\includegraphics[width=0.22\textwidth]{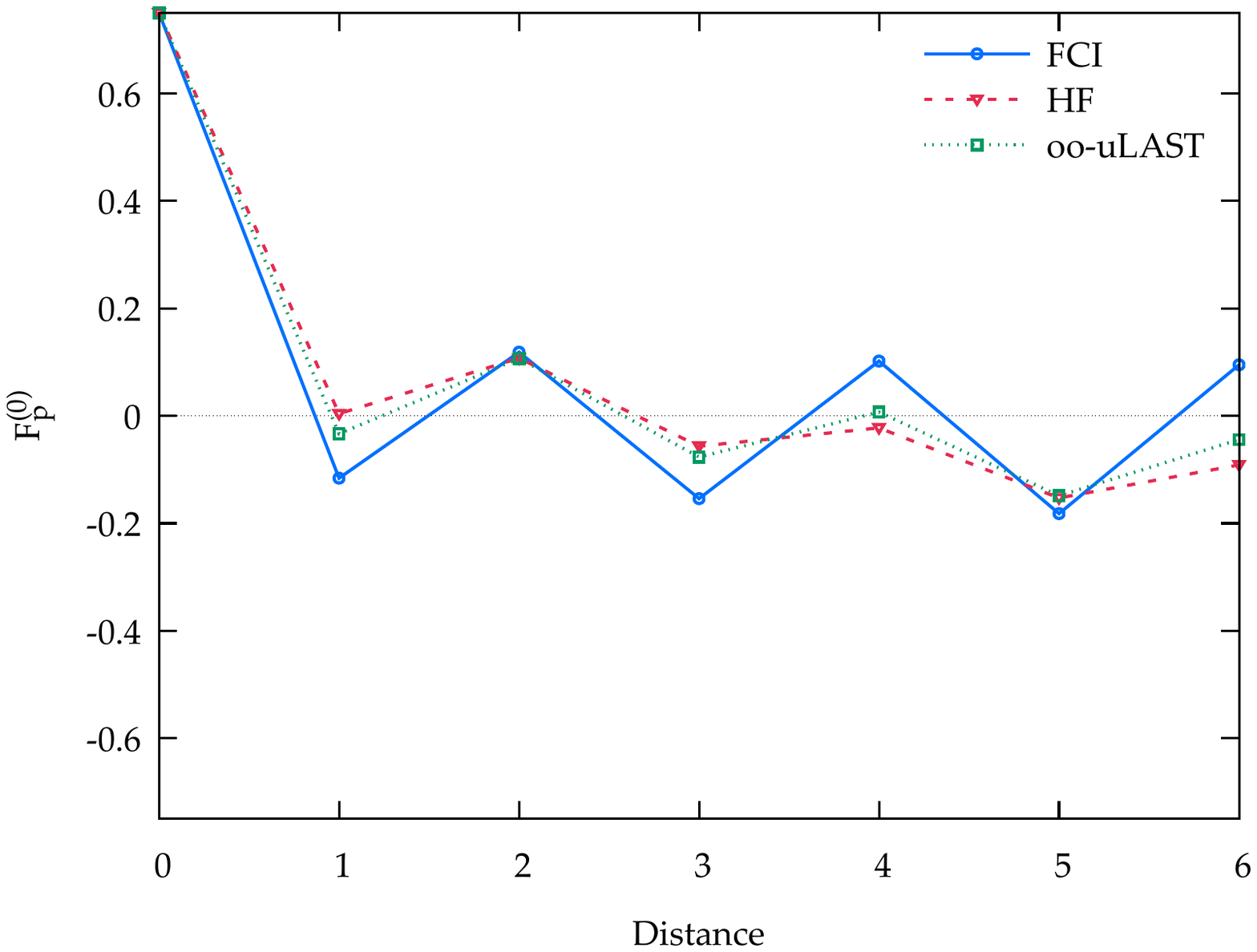}
\hfill
\includegraphics[width=0.22\textwidth]{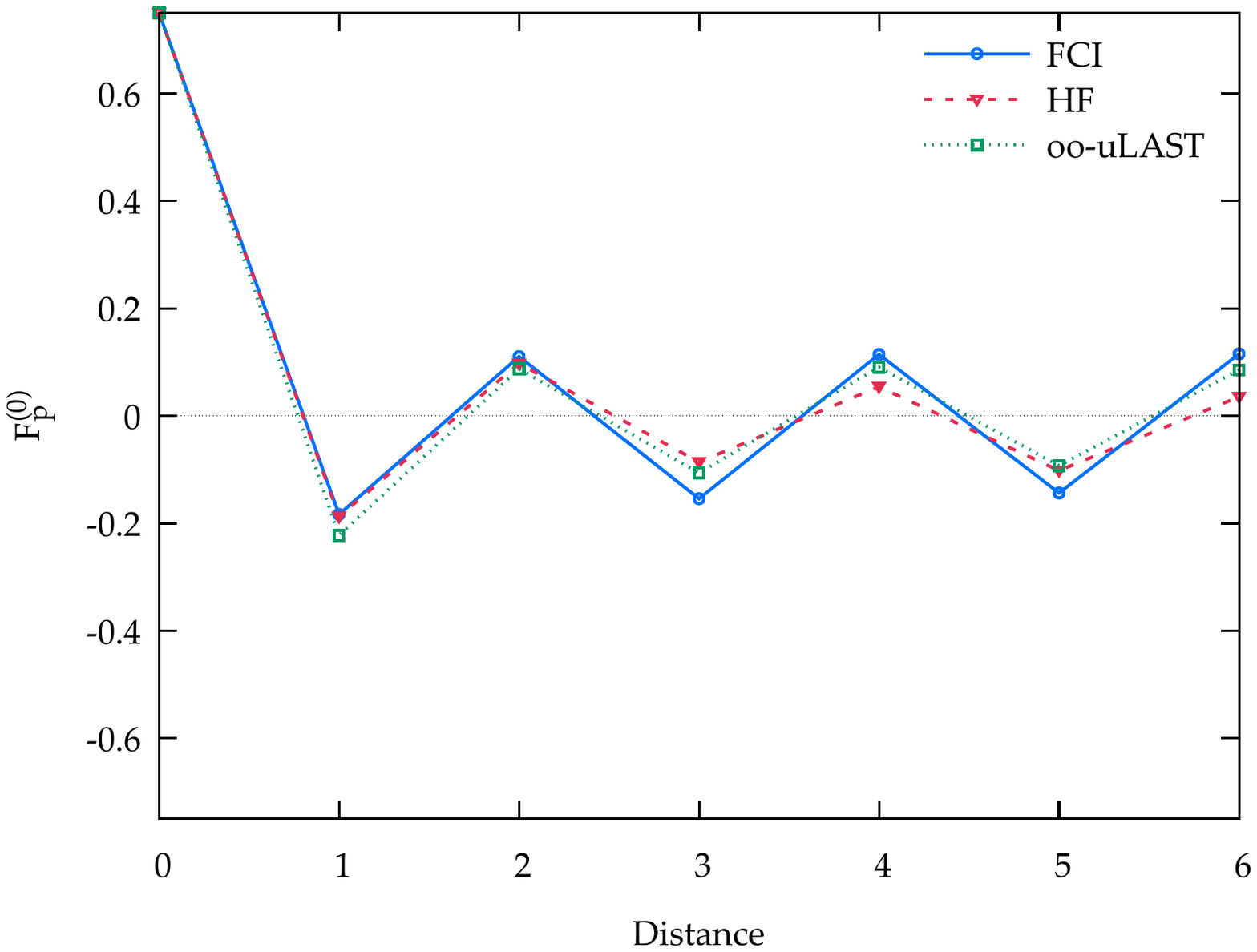}
\hfill
\includegraphics[width=0.22\textwidth]{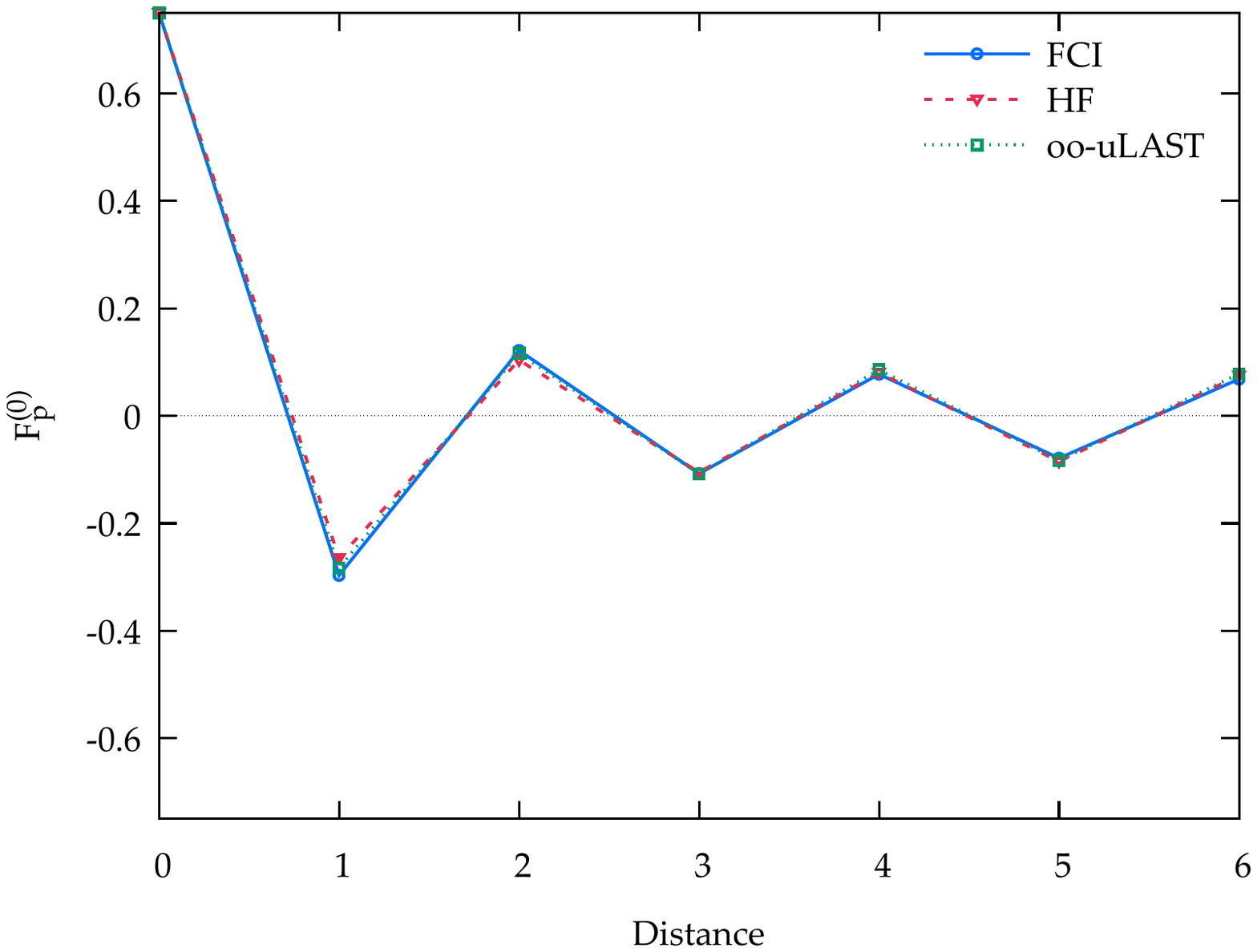}
\hfill
\includegraphics[width=0.22\textwidth]{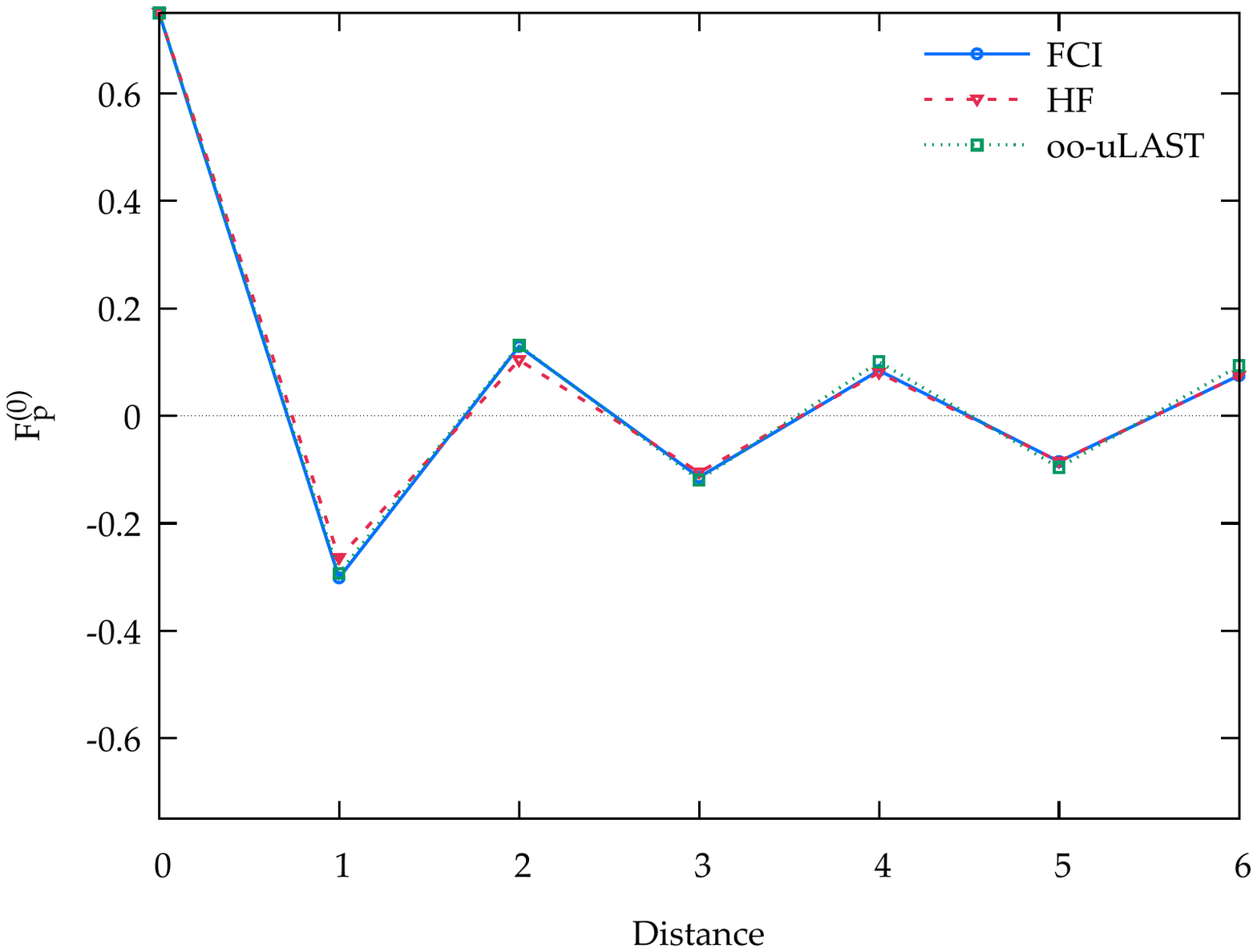}
\\
\includegraphics[width=0.22\textwidth]{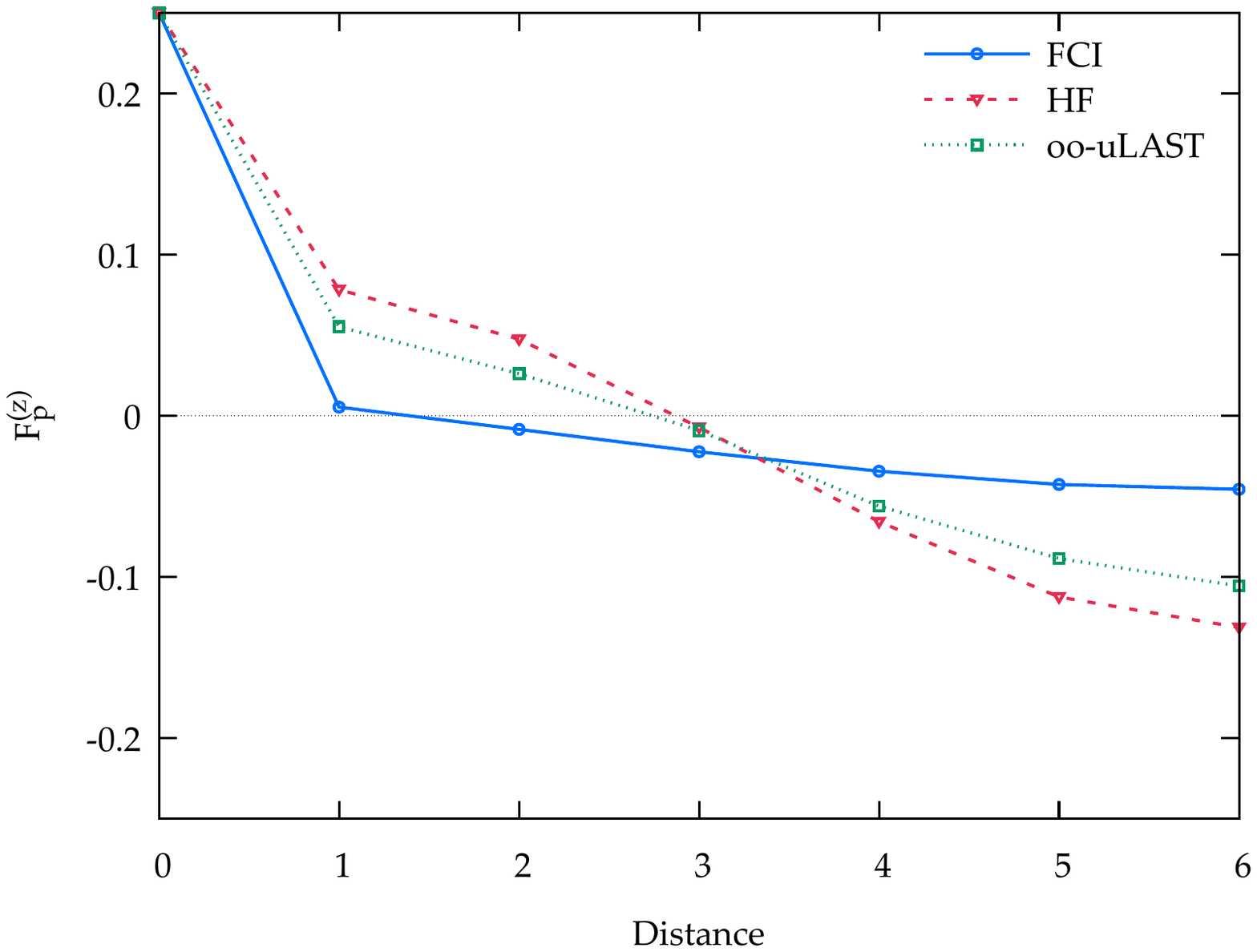}
\hfill
\includegraphics[width=0.22\textwidth]{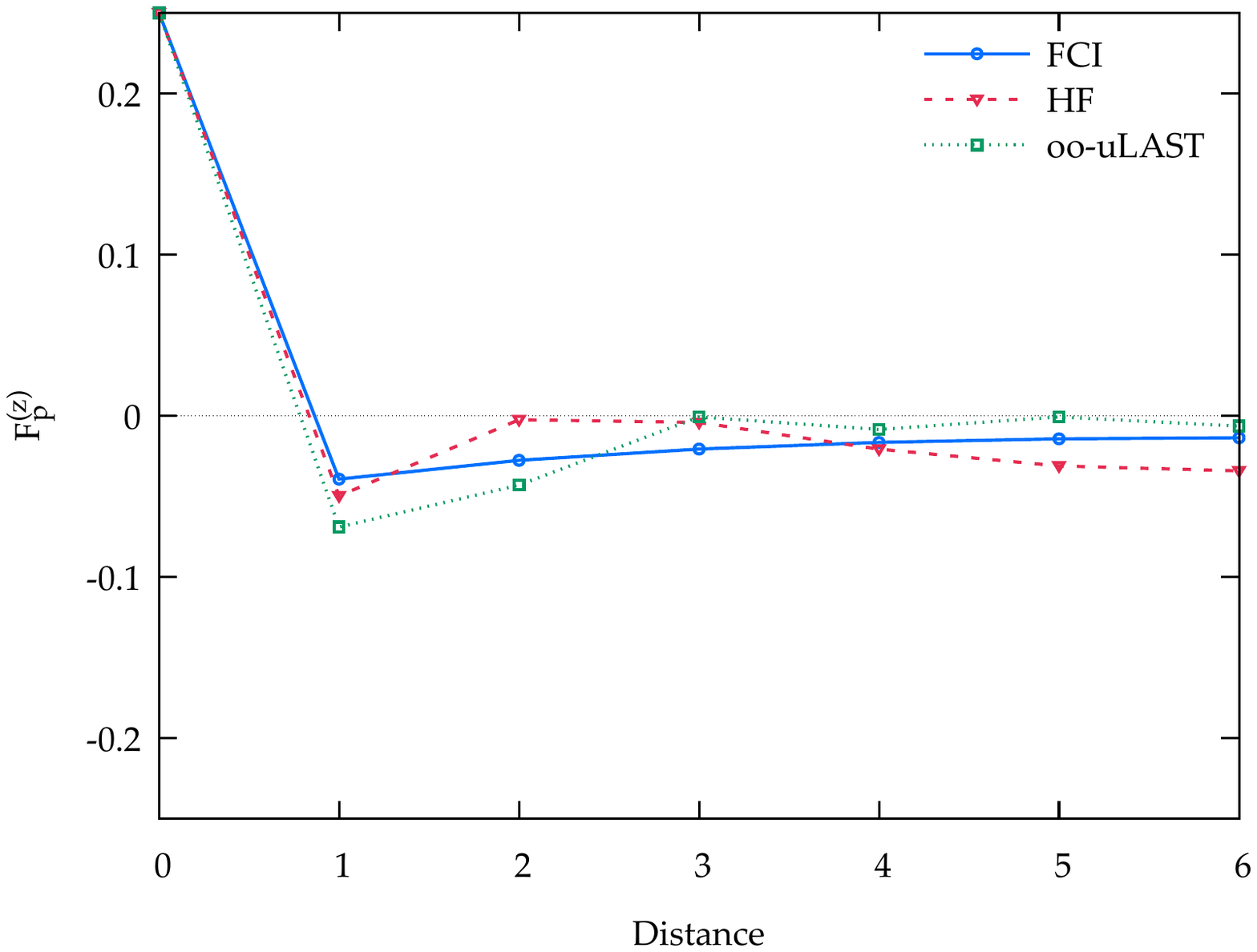}
\hfill
\includegraphics[width=0.22\textwidth]{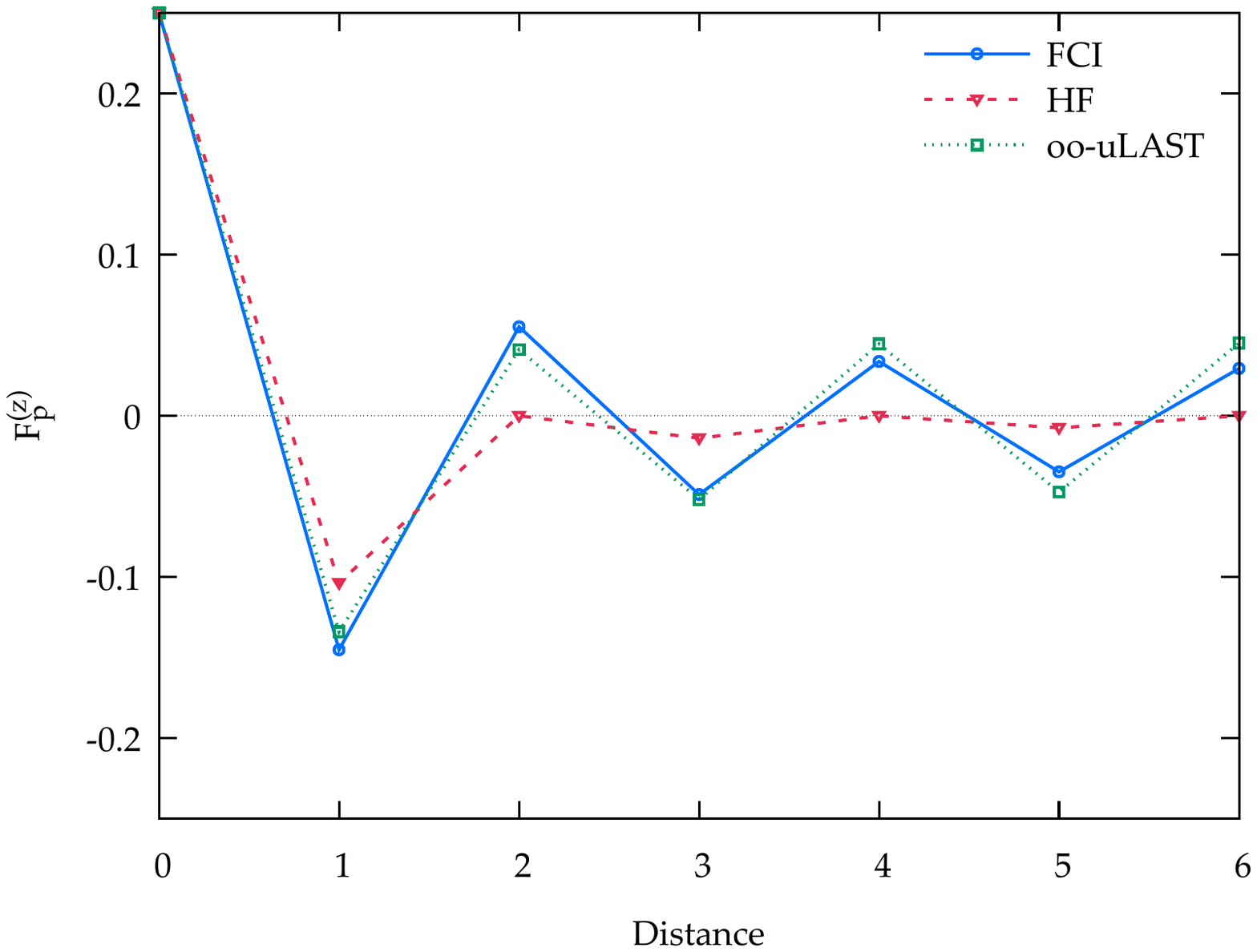}
\hfill
\includegraphics[width=0.22\textwidth]{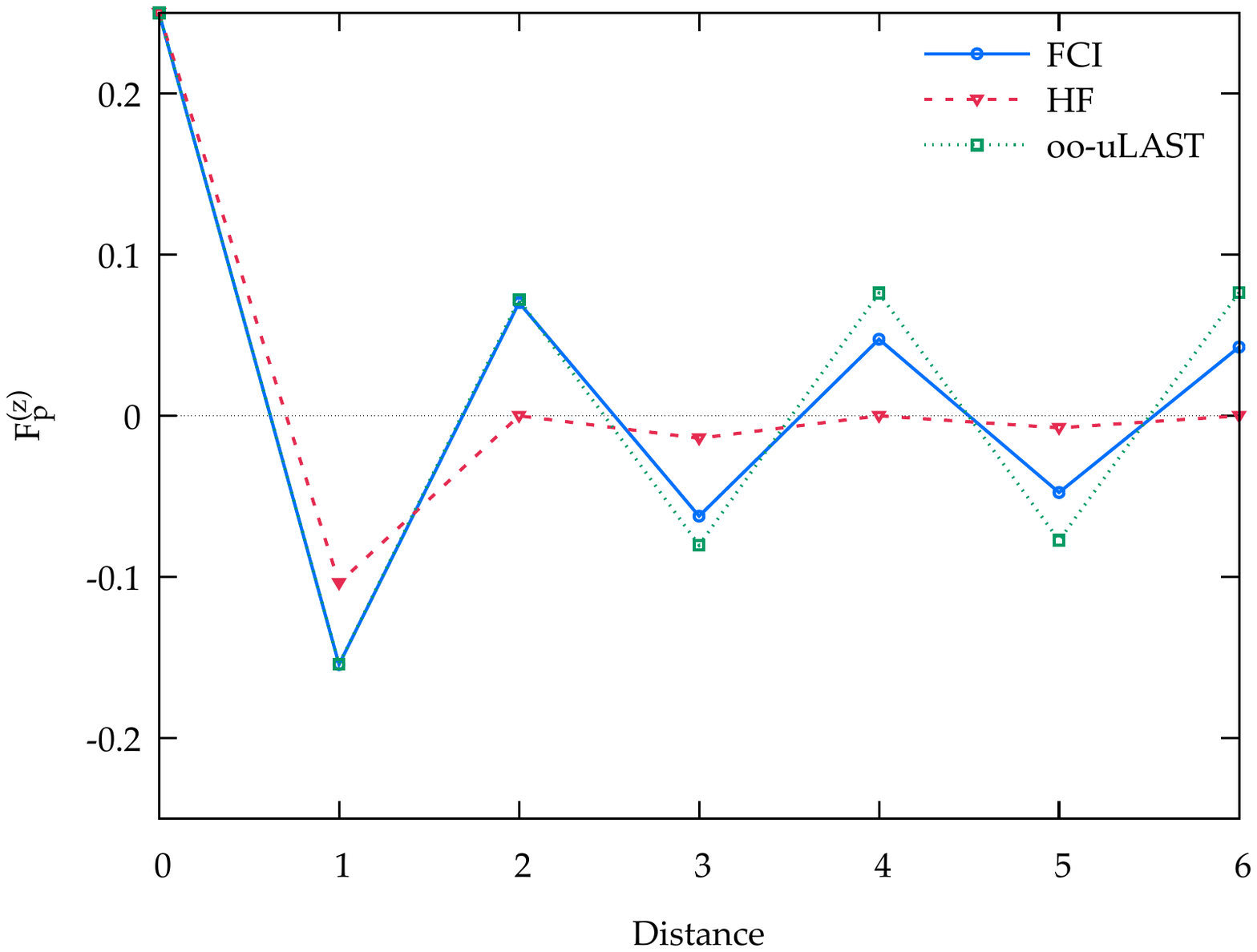}
\caption{Correlation functions in the 12-site periodic XXZ model in 1D with $S_z = 0$.  Top row: $\vec{S}\cdot \vec{S}$ correlation functions for $\Delta =  -1.1$, $\Delta = -0.9$, $\Delta = 0.9$, and $\Delta = 1.1$, respectively, from left to right.  Bottom row: $Sz-Sz$ correlation functions for these same values of $\Delta$.
\label{Fig:CorrFuncs}}
\end{figure*}

\subsection{Correlation Functions}
While total energies are generally the first thing we compute since they emerge naturally in the course of optimizing the wave function, they are not necessarily the most interesting.  A more relevant quantity in the context of spin systems are the spin-spin correlation functions.  We will confine ourselves to the 1D case with periodic boundary conditions, and define the $S^z-S^z$ and $\vec{S}\cdot\vec{S}$ correlation functions as, respectively,
\begin{subequations}
\begin{align}
F^{(z)}_p &= \sum_q \langle S^z_q \, S^z_{q+p} \rangle,
\\
F^{(0)}_p &= \sum_q \langle \vec{S}_q \cdot \vec{S}_{q+p}\rangle = F^{(z)}_p + \frac{1}{2} \, \sum_q \langle S^+_q \, S^-_{q+p} + S^-_q \, S^+_{q+p} \rangle,
\end{align}
\end{subequations}
where here we will take sites to be numbered in increasing order from one side of the lattice to the other.  After Jordan-Wigner transformation, these become
\begin{subequations}
\begin{align}
F^{(z)}_p &= \sum_q \langle \bar{n}_q \, \bar{n}_{q+p}\rangle,
\\
F^{(0)}_p &= F^{(z)}_p + \frac{1}{2} \, \sum_q \langle c_q^\dagger \, \phi_q^\dagger \, \phi_{q+p} \, c_p + c_q \, \phi_q \, \phi_{q+p}^\dagger \, c_{q+p}^\dagger\rangle,
\end{align}
\end{subequations}
where we recall that $\bar{n}_q = n_q - 1/2$ and where the expectation values are taken with respect to a Hartree-Fock or oo-uLAST wave function.  We do not include ST-LAST results in this section because the non-unitary similarity transformation in ST-LAST requires a biorthogonal expectation value as in traditional coupled cluster theory.

One important thing to note is that the $S^z-S^z$ correlation function is indepenent of the LAST parameters $\boldsymbol{\theta}$ since $\exp(\mathrm{i} \, \theta_2)$ commutes with the operator $\bar{n}_q \, \bar{n}_{q+p}$.  Thus, any differences between the $S^z-S^z$ correlation functions of HF and oo-uLAST are simply due to the differences between the $\boldsymbol{t}$ amplitudes defining the mean-field wave functions.

Figure \ref{Fig:CorrFuncs} shows correlation functions $F^{(0)}_p$ and $F^{(z)}_p$ for the 12-site XXZ model at four different values of $\Delta$.  We have picked these values as they bracket what become the phase transitions in the thermodynamic limit.  Generally, the $\vec{S}\cdot\vec{S}$ correlation functions from both HF and oo-uLAST are quite accurate (and of course are exact at $\Delta = 0$, where recall that HF is exact) except at $\Delta = -1.1$.  We note that the correlation functions improve as we move deeper into the ferromagnetic region $\Delta \lesssim -1$ (data not shown).  The $S_z-S_z$ correlation functions are of somewhat lower quality, though oo-uLAST offers a sigificant improvement over HF.  Since the $S_z-S_z$ correlation function is independent of $\boldsymbol{\theta}$, the improvements afforded by oo-uLAST simply reflect an underlying single determinant with a more physically correct structure.

\section{Conclusions}
The Jordan-Wigner transformation is a textbook example of a duality: certain spin-1/2 systems which are difficult to treat as spins can be readily handled after transformation to a system of spinless fermions, essentially because two-body spin operators like $S_p^+ \, S_q^-$ convert to one-body fermionic operators $c_p^\dagger \, c_q$ together with JW strings.  Because these strings are many-body operators and are frequently written in the form $\phi_p = \prod_{k<p} \left(1 - 2 \, n_k\right)$, the JW transformation has not been as widely exploited for the treatment of spin systems as it perhaps should be.  Yet while the JW strings are sufficiently cumbersome that they preclude a more sophisticated fermionic treatment, their action on fermionic determinants is simple, and we can treat the JW-transformed Hamiltonian with Hartree-Fock theory without undue difficulties.  Hartree-Fock-Bogoliubov or symmetry-projected mean-field methods\cite{Lowdin1955c,Ring1980,Blaizot1985,Schmid2004,JimenezHoyos2011b,JimenezHoyos2012}  are also straightforward.  While the symmetries of the JW-transformed Hamiltonian are generally complicated by the presence of strings, $S_z$ symmetry in the underlying spin Hamiltonian translates to number symmetry of the fermionic Hamiltonian, so number-projected Hartree-Fock-Bogoliubov,\cite{Ring1980,JimenezHoyos2011b} known in the chemistry community as the antisymmetrized geminal power,\cite{Coleman1965,Bratoz1965} could be applied readily.

On the other hand, these simple mean-field methods are not generally adequate for fermionic systems, and we cannot expect them to be adequate for general JW-transformed spin systems either.  While we cannot easily use traditional methods such as coupled cluster theory to correct the deficiencies of an HF treatment, our Lie algebraic similarity transformation theory is ideally suited to the task, because in practice its effect is simply to generalize the JW strings in providing a similarity-transformed Hamiltonian which is solved at the mean-field level.  In doing so, we should emphasize, we obtain extensive results (i.e. the LAST correlation energy and the Hartree-Fock mean-field energy both scale correctly with system size).  The combination of LAST and HF for JW-transformed systems offers excellent results at a quite reasonable computational cost.  Moreoever, this combination remedies perhaps the principle deficiency in the HF treatment of a JW-transformed Hamiltonian by providing results which are independent of the way in which we label sites.  We do not care to speculate about the implications of this invariance for other computational techniques which also require the user to specify a labeling of sites or orbitals, since it is, on first blush, simply due to the form the JW transformation.  But in the context of JW transformation, the improvements provided by LAST appear to be significant indeed.

Note that, while for these spin systems we obtain significantly better results by using fermionic mean-field methods on the JW-transformed Hamiltonian than by using equivalently costly spin-based methods, we do not imply that one should always prefer fermionic methods to spin-based approaches.  Indeed, some spin Hamiltonians can be exactly solved by a spin-based mean-field approach while their JW-transformed fermionic counterparts cannot be exactly solved by fermionic mean-field theories.\cite{Ryabinkin2018b}

Finally, our approach here of solving a Hamiltonian of spins by using the JW-transformation to convert it into a fermionic Hamiltonian stands as an interesting contrast to the program used in carrying out electronic structure calculations on a quantum computer, where the fermionic Hamiltonian is, by JW transformation, converted to a qubit (i.e. spin) form.\cite{Cao2019,Bauer2020} There is a spin-based counterpart to fermionic LAST,\cite{Khamoshi2021} and its unitary version might be of interest in this context.

\begin{acknowledgments}
This work was supported by the U.S. Department of Energy, Office of Basic Energy Sciences.  The Jordan-Wigner aspects were supported under Award DE-SC0019374, and the strong correlation aspects under Award DE-FG02-09ER16053.  G.E.S. is a Welch Foundation Chair (C-0036).
\end{acknowledgments}

\appendix
\section{Mathematical Details}
\subsection{Invariance of oo-uLAST
\label{Sec:AppendixGS}}
Recall that the extended Jordan-Wigner transformation writes
\begin{subequations}
\begin{align}
S_p^+ &\mapsto c_p^\dagger \, \phi_p^\dagger,
\\
S_p^z & \mapsto \bar{n}_p,
\\
\phi_p^\dagger &= \mathrm{e}^{\mathrm{i} \, \theta_{pq} \, n_q},
\end{align}
\end{subequations}
where $\theta_{pp} = 0$ and where
\begin{equation}
|\theta_{pq} - \theta_{qp}| = \pi.
\end{equation}
Conventionally, we might choose $\theta_{pq} = \theta_{qp} + \pi$ for $p > q$, but this choice is not essential.

A generic spin Hamiltonian
\begin{align}
H_s = \sum_p H_p \, S_p^z + \sum_{p \ne q} \Big[&W_{pq} \, S_p^z \, S_q^z
\\
 &+\frac{1}{2} \, V_{pq} \, \left(S_p^+ \, S_q^- + S_q^+ \, S_p^-\right)\Big],
\nonumber
\end{align}
becomes, upon this extended Jordan-Wigner transformation, a fermionic Hamiltonian
\begin{align}
H_f(\boldsymbol{\theta}) = \sum_p H_p \, \bar{n}_p &+ \sum_{p \ne q} \Big[W_{pq} \, \bar{n}_p \, \bar{n}_q
\\
 &+ \frac{1}{2} \, V_{pq} \, \left(c_p^\dagger \, \phi_p^\dagger \, \phi_q \, c_q + c_q^\dagger \, \phi_q^\dagger \, \phi_p \, c_p\right)\Big].
\nonumber
\end{align}

Now suppose we label the sites of that same spin Hamiltonian differently, transform it to a fermionic Hamiltonian as before, and then, after transformation, relabel the fermions to coincide with the labeling scheme of our original Hamiltonian.  For example, a 4-site Hamiltonian with sites ordered $1-3-2-4$ could be transformed to a fermionic Hamiltonian, after which the fermions can be relabeled in the order $1-2-3-4$.  This gives us two fermionic Hamiltonians with identical matrix elements $H_p$, $W_{pq}$, and $V_{pq}$, but different strings.  The string operators for the two different orders are each parameterized by a matrix $\boldsymbol{\theta}$ obeying the constraints we have already laid out.  Note, however, that the $\theta_{pq}$ are variational parameters to be optimized, each in the range $[-\pi,\pi)$ and, upon this optimization of the strings and mean-field wave function, we will obtain the same final result.

As we relabel, we may not have $\theta_{pq} = \theta_{qp} + \pi$ for $p > q$ but recall that factors of $2\pi$ are irrelevant so we are free to remove them.  For example, with sites ordered as $1-3-2-4$ we would have
\begin{equation}
\boldsymbol{\theta}^{(1)} = 
\begin{pmatrix}
0 & \theta_{13} & \theta_{12} & \theta_{14}
\\
\theta_{13} + \pi & 0 & \theta_{32} & \theta_{34}
\\
\theta_{12} + \pi & \theta_{32} + \pi & 0 & \theta_{24}
\\
\theta_{14} + \pi & \theta_{34} + \pi & \theta_{24} + \pi
\end{pmatrix}.
\end{equation}
Here, for convenience, we have labeled the rows and columns of the matrix $\boldsymbol{\theta}$ by the site to which they correspond.  Putting the fermions in the canonical order $1-2-3-4$ requires swapping the middle two rows and columns of the matrix, to
\begin{equation}
\boldsymbol{\theta}^{(2)} = 
\begin{pmatrix}
0 & \theta_{12} & \theta_{13} & \theta_{14}
\\
\theta_{12} + \pi & 0 & \theta_{32}+\pi & \theta_{24}
\\
\theta_{13} + \pi & \theta_{32} & 0 & \theta_{34}
\\
\theta_{14} + \pi & \theta_{24} + \pi & \theta_{34} + \pi
\end{pmatrix}.
\end{equation}
We can define $\theta_{32} = \theta_{23} + \pi$, which would give us
\begin{equation}
\boldsymbol{\theta}^{(2)} = 
\begin{pmatrix}
0 & \theta_{12} & \theta_{13} & \theta_{14}
\\
\theta_{12} + \pi & 0 & \theta_{23}+2\pi & \theta_{24}
\\
\theta_{13} + \pi & \theta_{23}+\pi & 0 & \theta_{34}
\\
\theta_{14} + \pi & \theta_{24} + \pi & \theta_{34} + \pi
\end{pmatrix}.
\end{equation}
Dropping the $2\pi$ gives us the same matrix $\boldsymbol{\theta}$ as we would have obtained had we directly labeled the sites in the order $1-2-3-4$, and of course we obtain the same result from these two orders upon variational optimization.

\subsection{Resumming the Similarity Transformation in LAST
\label{Sec:AppendixLAST}}
We have said that
\begin{equation*}
\left(c_p \, \alpha_2^n\right)_c =  c_p \, \alpha_{1,p}^n,
\end{equation*}
where we recall that
\begin{equation*}
\alpha_{1,p} = \sum_{q} \alpha_{pq} \, n_q
\end{equation*}
and that the notation $\left(c_p \, \alpha_2^n\right)_c$ means the $n$-tuple commutator of $\alpha_2$ with $c_p$.  Here, we wish to establish this inductively.

First, we note that it is trivially true for $n=0$, which just says that if we have no commutators of $\alpha_2$ with $c_p$, then the result is $c_p$.

We also note that definitionally,
\begin{equation}
\left(c_p \, \alpha_2^{n+1}\right)_c = [\left(c_p \, \alpha_2^n\right)_c,\alpha_2].
\end{equation}
Inserting our inductive hypothesis gives us
\begin{subequations}
\begin{align}
\left(c_p \, \alpha_2^{n+1}\right)_c
 &= [c_p \, \alpha_{1,p}^n,\alpha_2]
\\
 &=  \left(c_p \, [\alpha_{1,p}^n,\alpha_2] + [c_p^\dagger,\alpha_2] \, \alpha_{1,p}^n\right)
\end{align}
\end{subequations}
using the fact that $[AB,C] = A \, [B,C] + [A,C] \, B$.  Because $\alpha_{1,p}$ and $\alpha_2$ commute, we are left with
\begin{subequations}
\begin{align}
\left(c_p \, \alpha_2^{n+1}\right)_c
 &= \frac{1}{2} \, \sum_{rs} [c_p, n_r \, n_s] \, \alpha_{1,p}^n
\\
 &= \frac{1}{2} \, \sum_{rs} \alpha_{rs} \, \Big([c_p, n_r] \, n_s 
\\
 &\hspace{15mm} + n_r \, [c_p, n_s]\big) \, \alpha_{1,p}^n
\nonumber
\end{align}
\end{subequations}
after inserting the definition of $\alpha_2$.  Since $c_p$ and $n_r$ commute unless $p=r$, and $\alpha_{pq} = \alpha_{qp}$, we reduce this to
\begin{align}
\left(c_p \, \alpha_2^{n+1}\right)_c
 = \frac{1}{2} \, \sum_{q} \alpha_{pq} \, &\big([c_p, n_p] \, n_q
\\
&+ n_q \, [c_p, n_p]\big) \, \alpha_{1,p}^n.
\nonumber
\end{align}
Using 
\begin{equation}
[c_p, n_p] = c_p \, c_p^\dagger \, c_p - c_p^\dagger \, c_p \, c_p = c_p \, \left(1 - c_p \, c_p^\dagger\right) = c_p
\end{equation}
yields
\begin{equation}
\left(c_p \, \alpha_2^{n+1}\right)_c = \frac{1}{2} \, \sum_{q} \alpha_{pq} \, \left(c_p \, n_q + n_q \, c_p\right) \, \alpha_{1,p}^n.
\end{equation}
Finally, we note that $c_p$ and $n_q$ commute unless $p=q$ but that for $p=q$ the operator part is irrelevant because $\alpha_{pp} = 0$.  This gives us
\begin{equation}
\left(c_p \, \alpha_2^{n+1}\right)_c = c_p \, \left(\sum_{q} \alpha_{pq} \, n_q\right) \, \alpha_{1,p}^n = c_p \, \alpha_{1,p}^{n+1},
\end{equation}
completing the proof.

\bibliography{JordanWignerBib}

\end{document}